\documentclass[12pt]{article}
\usepackage{mathrsfs}
\usepackage{natbib,graphicx,setspace,lscape,longtable}
\usepackage{mathrsfs,amsmath,amsthm,amssymb,color}
\usepackage{natbib,epsfig,graphicx}
\usepackage{booktabs}
\usepackage{multirow}
\usepackage{hyperref}
\hypersetup{
     colorlinks   = true,
  linkcolor=black,
            urlcolor=black,
            citecolor=black
}

\bibpunct{(}{)}{;}{a}{,}{,}

\newtheorem{theorem}{Theorem}

\newtheorem{proposition}{Proposition}

\newcommand{\csection}[1]
    {\begin{center}
        \stepcounter{section}
        {\bf\large\arabic{section}. #1}
    \end{center}
}

\newcommand{\scsection}[1]
    {\begin{center}
        {\bf\large #1}
    \end{center}
}

\newcommand{\csubsection}[1]{
\begin{center}
\stepcounter{subsection}
{\it\arabic{section}.\arabic{subsection}. #1}
\end{center}
}

\def\beq{\begin{equation}}
\def\eeq{\end{equation}}
\def\beqr{\begin{eqnarray}}
\def\eeqr{\end{eqnarray}}
\def\beqrs{\begin{eqnarray*}}
\def\eeqrs{\end{eqnarray*}}
\def\bet{\begin{theorem}}
\def\eet{\end{theorem}}
\def\bel{\begin{lemma}}
\def\eel{\end{lemma}}
\def\bep{\begin{proposition}}
\def\eep{\end{proposition}}
\def\bg{\begin{figure}[tbph]\begin{center}}
\def\eg{\end{center}\end{figure}}

\def\bc{\begin{center}}
\def\ec{\end{center}}

\def\mQ{\mathcal Q}
\def\mR{\mathbb{R}}
\def\mS{\mathcal S}
\def\mX{\mathbb X}

\def\mY{\mathbb{Y}}

\def\be{\begin{equation}}
\def\ee{\end{equation}}
\def\ben{\begin{equation*}}
\def\een{\end{equation*}}
\def\bea{\begin{eqnarray}}
\def\eea{\end{eqnarray}}

\def\bda{\begin{eqnarray*}}
\def\eda{\end{eqnarray*}}

\def\laak{\left \| }
\def\raak{\right \| }

\def\be{\begin{equation}}
\def\ee{\end{equation}}
\def\ben{\begin{equation*}}
\def\een{\end{equation*}}
\def\bea{\begin{eqnarray}}
\def\eea{\end{eqnarray}}

\def\bda{\begin{eqnarray*}}
\def\eda{\end{eqnarray*}}

\def\laak{\left \| }
\def\raak{\right \| }

\numberwithin{equation}{section}

\begin{document}

\begin{center}
{\bf\Large Imputations for High Missing Rate Data in Covariates via  Semi-supervised Learning Approach}\\
\bigskip
Wei Lan, Xuerong Chen, Tao Zou and Chih-Ling Tsai \\
{\it Southwestern University of Finance and Economics, The Australian National University and University of California, Davis}\\

\end{center}

\begin{abstract}
Advancements in data collection techniques  and the heterogeneity of data resources can
yield high percentages of missing observations on  variables, such as block-wise missing data. Under missing-data scenarios, traditional methods such as the simple average, $k$-nearest neighbor, multiple, and regression imputations may lead to  results that are unstable or unable be computed.
Motivated by the concept of semi-supervised
learning  (see, e.g., Zhu and Goldberg, 2009 and Chapelle et al., 2010), we propose a novel approach
with which to fill in missing values in covariates that have high missing rates.
Specifically, we consider the missing and non-missing  subjects in any covariate as the unlabelled and labelled target outputs, respectively,
and treat their corresponding responses as the unlabelled and labelled inputs. This innovative setting allows us to
impute a large number  of missing data without imposing any model assumptions. In addition, the resulting imputation has a closed form for continuous covariates,
and it can be calculated efficiently. An analogous procedure is applicable for discrete covariates. We further
employ the nonparametric techniques to show the theoretical properties of imputed covariates.
Simulation studies and an online consumer finance example are presented
to illustrate the usefulness of the proposed method.

\bigskip
\noindent {\bf{KEY WORDS}}: Block-wise Missing; Cross-validation; High Missing Rate Data;
Interchangeable Imputation; Semi-supervised Imputation

\end{abstract}

\newpage

\csection{INTRODUCTION}

Missing data is a common challenge encountered by researchers and practitioners.
This is because ignoring
missing values often
results in loss of information and
yields biased parameter estimates.
To account for possible bias and retain the representativeness of the data, several
traditional techniques have been proposed such as
 $k$-nearest neighbor (KNN, Troyanskaya et al., 2001), regression imputation (RI, Little and Rubin, 2002; Shao and Wang, 2002), multiple imputation
(MI, Little and Rubin, 2002), random forest (RF, Stekhoven and B\"uhlmann, 2012),
complete-case analysis (CC, see Little, 1992), full-information maximum likelihood estimation (FIME, Collins, et al., 2001; Myrtveit et al., 2001; Allison, 2012),
weighted estimation equation (WEE, Robins et al., 1994; Lipsitz et al., 1999) and multiple robust estimation (MRE, Han and Wang, 2013; Chen and Haziza, 2017).

Among the aforementioned methods, FIME, WEE, and MRE mainly aim to alleviate bias and improve efficiency in parameter estimations. Hence, they often
require model structure and additional assumptions.
On the other hand,  KNN, RF, MI and RI
generally focus on imputation of the missing observations.
These methods usually do not need to impose any model structure nor other assumptions.
Although the popular CC method does not necessarily require imposing assumptions, it is not used for imputation.
None of these methods are  designed for addressing high missing rate data in covariates.
Accordingly, their imputations can be unstable or even uncomputable.

Due to the rapid development of advanced technology and the heterogeneity of data resources, it is expected that data collection with high missing rates will become more prevalent.
One  example is  data  collected from multiple sources in which
only a small portion of observations contain the complete information across all sources.
In this case, the data has block-wise missing entries, which can result in high missing rates
in covariates (see, e.g., Xiang et al., 2014 and Fang et al., 2019).
Another example is data collected from online survey in which the respondents are not required to
answer all questions. As a result, some questions, such as income and age, can have low response rates due to privacy, and thus yields high missing rates (see, e.g., Nulty, 2008;
Bollinger and Hirsch, 2013).
This paper considers a real application for risk management in online consumer finance in which we  assess the repayment ability of each loan applicant in order to determine their credit quota.
Table 1 presents the data
collected from five sources,  yielding six distinct patterns of high missing rates;
the total missing rate is 49.4\%, and
only 4.7\% applicants have complete covariate
information. A detailed description of this example can be found in Section 3.

Inspired by  practical need, there are several methods proposed in the literature that manage
high missing rate data such as block-wise missing data (see, e.g., Zhou et al., 2010, Yuan et al. 2012 and Xiang et al., 2014).
These methods basically require some model assumptions in order to achieve their specific goals.
To complement those approaches,
we propose a new procedure
motivated by semi-supervised learning (SSL; see, e.g., Zhu and Goldberg, 2009
and Chapelle et al., 2010).
Specifically, we consider the missing and non-missing subjects in any covariate
as the unlabelled and labelled target outputs, respectively.
Then we treat their corresponding responses as  the unlabelled and labelled inputs.
Under this innovative setting, we are able to develop an imputation method
that is applicable for high missing rate data
without imposing any model structure assumptions.
Since our proposed method derives from semi-supervised learning, we simply name it semi-supervised imputation (SSI).

The SSI method utilizes information from both observed and unobserved subjects across responses and covariates.
Suppose that the $i$-th subject is missing at the $j$-th covariate $\mX_j=(X_{1j},\cdots, X_{nj})^\top\in\mR^n$   (i.e., $X_{ij})$.
We then formulate it
as a function of the weighted
average of the rest of the subjects in $\mX_j$.  The weight between $X_{ij}$ and any other
subject $X_{lj}$ ($l\ne i$) is determined by their
similarity (or distance) measure, which is constructed by using the $X_{ij}$'s and $X_{lj}$'s  corresponding responses and their commonly observed covariates.
We also allow the weight to depend on scale-parameters for controlling the magnitude of similarity.  As a result, we are able to obtain a closed-form
imputation of $X_{ij}$.

Formulating the imputation task through the concept of semi-supervised learning has three important merits.
First,  SSL has the capability of handling large amounts of unlabeled data with promising
results (e.g., see Zhu et al., 2003).
Accordingly,
SSI enables us to manage high missing rate data with a model-free process regardless of missing mechanisms such as missing completely at random,
missing at random, and missing not at random.
Second, SSI has the closed form for continuous covariates, which can be calculated efficiently even for a large number of missing subjects. In addition, this closed form allows us to develop theoretical properties  of SSI. An analogous approach is applicable to discrete covariates.
Third, SSI not only uses the information from the observed responses and covariates,
but also utilizes the information from those covariates with missing subjects. Thus, SSI can yield more reliable results than commonly used methods such as KNN, RF, MI, and RI.

The rest of this article is organized as follows. Section 2 introduces SSI
and demonstrates the theoretical properties of imputed values with continuous covariates and discrete covariates, respectively.
In addition, the interchangeable imputation method  for finding unknown scale-parameters is proposed. Based on both observed data and imputed values, we further introduce a linear regression model. This  allows us not only to obtain
regression parameter estimators and predictions, but also to propose a cross-validation approach for estimating unknown scale-parameters.
Monte Carlo studies and an online consumer finance example are presented in Section 3, which  indicate that SSI performs well. The article
 concludes with some discussion in Section 4, while
 all technical details are relegated to the supplementary material.

\csection{SEMI-SUPERVISED IMPUTATION FOR COVARIATES}

This section contains three subsections, namely SSI for continuous covariates, SSI for discrete covariates, and scale-parameters selection.

\csubsection{SSI for continuous covariates}

Consider a data set $\{(Y_i,X_{ij}),i=1,\cdots,n,j=1,\cdots,p\}$ containing $n$ subjects and $p$ covariates.
Let $\mY=(Y_1,\cdots, Y_n)^\top\in\mR^n$ be the response vector. We assume it to be continuous and fully observed throughout the entire article, while
the discussions for partially observed $\mY$ are given in Section 4. Denote
the $i$-th subject of the $p$ covariates $X_i=(X_{i1}, \cdots, X_{ip})^\top\in\mR^p$
and the $j$-th covariate for the $n$ subjects  $\mX_j=(X_{1j}, \cdots, X_{nj})^\top\in\mR^n$ as defined in Section 1.
We also assume that all $X_i$s are continuous and some of the $X_{ij}$s are missing.
Moreover, define $D_{ij}=0$ if covariate $X_{ij}$ is missing, and $D_{ij}=1$
otherwise.

We next adapt semi-supervised learning to fill in a larger number of missing values.
Specifically, for every covariate $\mX_j$
($j=1, \cdots, p$),
we treat its  missing and observed subjects as  unlabeled and labeled target outputs, respectively, while we regard the corresponding responses in $\mY$ as  unlabeled and labeled inputs.
We then model the $j$-th covariate of the $i$-th subject,
$X_{ij}$, as a weighted average of the remainder of the subjects in $\mX_j$, where the
weights are determined by similarity measures (or distances)
between the $i$-th and any other subjects (see, e.g., Zhu et al., 2003; Belkin et al., 2006).

In general, if the two subjects' corresponding responses and covariates are similar, their
distance should be small. Hence, for any two
different subjects $i_1$ and $i_2$, we
construct the weight matrix $A=(a_{i_1i_2})\in\mR^{n\times n}$  given below (see, e.g., Zou et al., 2017).
\beq
  a_{i_1i_2}=K_{h_1}(Y_{i_1}-Y_{i_2})\prod_{k\in D_{i_1}\cap D_{i_2}}K_{h_2}(X_{i_1k}-X_{i_2k}),
\eeq
where $K_{h_l}(\cdot)=K(\cdot/h_l)$, $K(\cdot)$ is a kernel function,  $h_l$ for $l=1$ and $2$ are
bandwidth parameters, and $D_i=\{j: D_{ij}=1\}$ contains the observed covariates at subject $i$.
When $\mX_k$ is discrete, we define $K_{h_2}(X_{i_1k}-X_{i_2k})=K_{h_2}(0)$ if $X_{i_1k}$
and $X_{i_2k}$ belong to the same category and $K_{h_2}(X_{i_1k}-X_{i_2k})=K_{h_2}(1)$ otherwise (see, e.g., Rojagopalan and Lall, 1995).
In this paper, we consider  the popular kernel function used in machine learning, the radial basis function kernel
 (i.e., Gaussian kernel). Accordingly,
 the resulting weight is
\[a_{i_1i_2}=\exp\Big\{-\lambda_1(Y_{i_1}-Y_{i_2})^2-\lambda_2\sum_{k\in D_{i_1}\cap D_{i_2}}(X_{i_1k}-X_{i_2k})^2\Big\},\]
where $\lambda=(\lambda_1, \lambda_2)^\top$, $\lambda_j=h_j^{-1}$ ($j=1,2$) are unknown scale parameters that need to be determined from the data.

Based on the weight matrix $A$, we then propose a weighted average approach in order to impute
missing observations in covariates. For any $j=1,\cdots,p$,
define  $\mS_{0j}=\{i: D_{ij}=0\}$ and $\mS_{1j}=\{i: D_{ij}=1\}$, thus they represent the missing and non-missing subjects in covariate $j$, respectively.
Accordingly, $X_{\mS_{0j}j}=(X_{ij}: i\in\mS_{0j})$ and $X_{\mS_{1j}j}=(X_{ij}: i\in\mS_{1j})$ are  the missing and non-missing subsets of $\mX_j$.
Adopting the concept of semi-supervised learning  in Zhu et al. (2003) and Wasserman (2007), we consider that
the missing observation $X_{ij}$ for $i\in\mS_{0j}$ is closer to those observations that are nearer to subject $i$ as  induced by their associated weights in $A$.
This motivates us to
impute missing observations $X_{\mS_{0j}j}$ by
minimizing the following quadratic loss function with respect to $X_{ij}^{*}$  under the constraint $X_{ij}^{*}=X_{ij}$ for the observed subjects
(see, e.g., Zhu et al., 2003; Zhang and Li, 2006; Belkin et al., 2006; Du and Zhao, 2017):
\beq \min_{X_{ij}^{*}, i=1, \cdots, n} \sum_{i_1=1}^n \sum_{i_2=1}^n w_{i_1i_2} (X^*_{i_1j}-X^*_{i_2j})^2, \mbox{~subject to~} X^*_{ij}=X_{ij} \mbox{~for any~} i\in\mS_{1j}, \eeq
where $w_{i_1i_2}=a_{i_1i_2}/\sum_{i_2} a_{i_1i_2}$ are the row-normalized versions of $a_{i_1i_2}$ and they constitute the weighted matrix
$W=(w_{i_1i_2})$.

To solve (2.2), we set the first derivative of the objective function (2.2) with respect to $X_{i_1j}^*$, for any $i_1\in\mS_{0j}$, to be 0. As a result,
 $\sum_{i_2=1}^n w_{i_1i_2} (X_{i_1j}^*-X_{i_2j}^*)=0$ for any $i_1\in\mS_{0j}$.
Note that $W$ is a row-normalized matrix and $\sum_{i_2=1}^n w_{i_1i_2}=1$. Accordingly, we  obtain the following imputation algorithm for every subject $i_1\in\mS_{0j}$,
\beq
X_{i_1j}^*=\sum_{i_2\in\mS_{0j}} w_{i_1i_2} X_{i_2j}^*+\sum_{i_2\in\mS_{1j}} w_{i_1i_2} X_{i_2j}.
\eeq
Define $C_i=\sum_{i^*\in\mS_{1j}} w_{ii^*} X_{i^*j}$, $C_{\mS_{0j}}=(C_i: i\in\mS_{0j})$, and
$W_{\mS_{0j}}=(w_{i_1i_2}: i_1\in\mS_{0j}, i_2\in\mS_{0j})$.
Then Equation (2.3) can be  expressed as
\beq X_{\mS_{0j}j}^*=W_{\mS_{0j}}X_{\mS_{0j}j}^*+C_{\mS_{0j}},\eeq
which has the closed form solution $X_{\mS_{0j}j}^*=(I-W_{\mS_{0j}})^{-1}C_{\mS_{0j}}$, and
$I-W_{\mS_{0j}}$ is invertible by
Condition (C5) (see, Du and Zhao, 2017).
Hence, for $j=1,\cdots, p$, one can impute the missing subjects in $X_{\mS_{0j}j}$
without imposing any model assumptions.
Furthermore,
define $W_{\mS_{0j}\mS_{1j}}=(w_{i_1i_2}, i_1\in\mS_{0j}, i_2\in\mS_{1j})$.
As a result,
 $C_{\mS_{0j}}=W_{\mS_{0j}\mS_{1j}}X_{\mS_{1j}j}$,
and according to (2.4), the missing subjects of $X_{\mS_{0j}j}$
can be imputed by
\beq
\hat X_{\mS_{0j}j}=(I-W_{\mS_{0j}})^{-1}W_{\mS_{0j}\mS_{1j}}X_{\mS_{1j}j},
\eeq
which is a weighted average of the observed  subjects in $X_{\mS_{1j}j}$.
Since the above imputation procedure is  motivated from semi-supervised learning, we denote our proposed approach SSI as mentioned in Section 1.
It is worth noting that SSI can yield
reliable results in the imputation of a large number of missing subjects since it takes into account all information from both responses and covariates that
are related to missing subjects. Simulation results in Section 3 support this finding.

For continuous covariates, we investigate the average performance of $\hat X_{\mS_{0j}j}$ given below.

\bet
Assume that Conditions (C1)-(C6) hold in Appendix A of the supplementary material.
Then, for any given $\delta>0$, there exist finite positive constants $C_1$ and $C_2$ such that,
for each $j=1, \cdots, p$,
 \[
 P\bigg(\bigg|\frac{1}{|\mS_{0j}|}\sum_{i\in \mS_{0j}}(\hat{X}_{ij}-X_{ij})\bigg|>\frac{\delta}{n}\bigg)\leq 6\exp\bigg(-\frac{1}{2}\frac{\delta^2}{C_1n+C_2\delta}\bigg).
 \]
\eet

\noindent
The above theorem implies that
$|\mS_{0j}|^{-1}\sum_{i\in \mS_{0j}}(\hat{X}_{ij}-X_{ij})\rightarrow_p 0$ as $n\rightarrow\infty$ under Conditions (C1)-(C6). Thus,
the sample average of the imputed values via SSI
 consistently estimates the true average of the missing values.

\noindent\textbf{Remark 1:} To construct the weight matrix, we employ the widely used Gaussian kernel
to measure the distance between any two different subjects. However, this does not exclude other possible kernel functions such as polynomial kernels. In comparing with that kernel, there are two major reasons for using the Gaussian kernel.
First, the Gaussian kernel
is equivalent to mapping the data into an infinitely differentiable function space. As a result, it can partially capture the properties of
polynomial kernels with high polynomial degrees. Second, the Gaussian kernel has fewer tuning parameters to be calculated, so  it can be computationally efficient
in large data sets.
Our simulation studies and real data analyses indicate that the Gaussian kernel performs satisfactorily.

In contrast to continuous covariates, we next study discrete covariates with missing subjects.

\csubsection{SSI for discrete covariates}

Suppose that $\mX_j$ is discrete and has $C$ class labels $\{1, \cdots, C\}$. We then define label matrix $\bar \mX_j=(\bar X_{ic,j})\in\mR^{n\times C}$,
where $\bar X_{ic, j}$ denotes the probability of the $j$-th covariate and subject $i$
in class $c$.
Note that if $X_{\mS_{1j}j}$ is observed, then, for any $i\in\mS_{1j}$, we have $\bar X_{ic, j}=1$
if $X_{ij}=c$, and $\bar X_{ic, j}=0$ otherwise.
In addition, define
$\bar X_{\mS_{0j}j}=(\bar X_{ic, j},i\in\mS_{0j}, 1\leq c\leq C)$ and $\bar X_{\mS_{1j}j}=(\bar X_{ic, j},i\in\mS_{1j}, 1\leq c\leq C)$.
We then adopt Equation (2.4) and propose the following algorithm, which enables us to impute missing subjects efficiently.
\begin{itemize}
\item [(1.)] Set an initial value of $\bar X_{\mS_{0j}j}$, and name it as $\bar X_{\mS_{0j}j}^{(0)}$. Then update $\bar X_{\mS_{0j}j}$
by $\bar X_{\mS_{0j}j}^{(1)}=W_{\mS_{0j}}\bar X_{\mS_{0j}j}^{(0)}+\bar C_{\mS_{0j}}$ with $\bar C_{\mS_{0j}}=W_{\mS_{0j}\mS_{1j}}\bar X_{\mS_{1j}j}$. Subsequently normalize the rows of $\bar X_{\mS_{0j}j}^{(1)}$.
\item [(2.)] In the $k$-th step, for given $\bar X_{\mS_{0j}j}^{(k)}$, update $\bar X_{\mS_{0j}j}$
by $\bar X_{\mS_{0j}j}^{(k+1)}=W_{\mS_{0j}}\bar X_{\mS_{0j}j}^{(k)}+\bar C_{\mS_{0j}}$, and normalize the rows of $\bar X_{\mS_{0j}j}^{(k+1)}$.
\item [(3.)] Stop at the $K$-th step if $\|\bar X_{\mS_{0j}j}^{(K)}-\bar X_{\mS_{0j}j}^{(K-1)}\|<\epsilon$ for some pre-specified value $\epsilon$,
where $\laak \cdot\raak$ denotes the Euclidean norm of any arbitrary vector. Afterwards,  normalize the rows of $\bar X_{\mS_{0j}j}^{(K)}$. Finally,
for any $i\in\mS_{0j}$, obtain $\hat X_{ij}=\mbox{argmax}_{1\leq c\leq C} \bar X^{(K)}_{ic,j}$.
\end{itemize}

The above algorithm generates imputations for discrete covariates. Note that, for each iteration, the algorithm has the
closed form. Thus, it is convergent regardless of the initial value under Condition (C5) (see the discussion of Zhu et al., 2003).
We next investigate the
theoretical property of this algorithm.

\bet
Assume that Conditions (C1)-(C6) hold in Appendix A of the supplementary material.
Then, for any given $\bar\delta>0$, there exist finite positive constants $\bar C_1$ and $\bar C_2$ such that, for each $j=1, \cdots, p$,
 \[
 P\bigg(\bigg|\frac{1}{|\mS_{0j}|}\sum_{i\in \mS_{0j}}(\bar X^{(K)}_{ic,j}-\bar X_{ic,j})\bigg|>\frac{\bar\delta}{n}\bigg)\leq 6\exp\bigg(-\frac{1}{2}\frac{\bar\delta^{2}}{\bar C_1n+\bar C_2\bar\delta}\bigg),
 \]
where $K$ is defined in the step (3) of the above algorithm.
\eet

\noindent
The proof of Theorem 2 is quite similar to that of Theorem 1; we thus omit it to
save space.
This theorem shows that, for each of the discrete covariate,
the sample average of the probability of imputed values belonging to class $c$ via SSI
consistently estimates the true average probability of missing values.

\noindent\textbf{Remark 2:}
In the missing data  with mixing continuous and discrete covariates, one can construct the weight matrix by integrating the weights from both continuous covariates  and discrete covariates via (2.1) and the formula under (2.1), respectively.
Afterwards,
one can employ (2.5) and   the algorithm proposed in Section 2.2 to impute missing values for continuous and discrete covariates, separately.

\csubsection{Scale-parameters Selection, Estimation and Prediction}

In semi-supervised learning, one can assume that the scale parameter in the weight matrix is known
(see, e.g., Zhu and Goldberg, 2009). Based on prior knowledge and previous experience,
 practitioners can take the same approach to  implement SSI by using  the two known scale parameters.
However, this  approach is subjective, which motivates us to propose a data-driven and model-free procedure.
According to Condition (C2), we recommend using $\lambda=O(n^{1/(2d_0+1)})$, where $d_0=1+\max_{1\leq i\leq n-1}|D_{i}\cap D_{i+1}|$ defined in Appendix A of the supplementary material. Let
$\tau=\lambda n^{-1/(2d_0+1)}$, and then employ the interchangeable imputation method given below to select the
optimal $\tau$.

For the $j$-th covariate with given $\tau$, we apply SSI to impute $X_{\mS_{0j}j}$ and obtain $\hat X^{\tau}_{\mS_{0j}j}$.
Based on the imputed values of $\hat X^{\tau}_{\mS_{0j}j}$, we interchangeably treat $X_{\mS_{1j}j}$ as missing and employ SSI again to yield $\hat X^{\tau}_{\mS_{1j}j}$. This procedure allows us to
compute the imputation error of $\hat X^{\tau}_{\mS_{1j}j}$, which is $\|\hat X^{\tau}_{\mS_{1j}j}-X_{\mS_{1j}j}\|^2$.
Repeat the same procedure for $j=1, \cdots, p$, and obtain the overall imputation error,
$\mQ(\tau)=\sum_{j=1}^p \|\hat X^{\tau}_{\mS_{1j}j}-X_{\mS_{1j}j}\|^2$. Finally, we
select $\tau$ as $\hat\tau=\mbox{argmin}_{\tau} \mQ(\tau)$ by minimizing the overall imputation error  with a grid-search over the pre-specified bounded region.
As a result, the selected $\hat\lambda=\hat\tau n^{1/(2d_0+1)}$
should be of order $O(n^{1/(2d_0+1)})$.

After selecting scale parameters, one can not only impute missing values, but also
conduct subsequent data analysis such as estimation, prediction, and classification. Since the focus of our paper
is prediction, we first assume that there exists a relationship between the responses and the imputed-observed covariates via the linear regression model, $Y_i=\hat X_i^{\top}(\hat\tau)\beta+\epsilon_i$,
where $\hat X_i(\hat\tau)=(\hat X_{i1}(\hat\tau),\cdots,\hat X_{ip}(\hat\tau))^{\top}$,
$\hat X_{ij}(\hat\tau)=X_{ij}$ if the $i$-th subject in $j$-th covariate is observed for
$i=1, \cdots, n$ and  $j=1,\cdots,p$, and
$\beta=(\beta_1,\cdots, \beta_p)^\top$
is the $p\times 1$ vector of unknown regression coefficients. In addition, $\epsilon_i$s are random errors
with mean 0 and variance $\sigma^2$.
Then, the resulting ordinary least squares estimator of $\beta$ is
$\hat\beta(\hat\tau)=(\hat \mX^{\top}(\hat\tau)\hat \mX(\hat\tau))^{-1}\hat\mX^{\top}(\hat\tau) \mY$,
where  $\mX(\hat\tau)=(\hat X_1(\hat\tau),\cdots,\hat X_n(\hat\tau))^{\top}$.
Then,  for any
given $\hat X_{0}(\hat\tau)$, we obtain the prediction $\hat Y_0=\hat X_{0}({\hat\tau})^\top\hat\beta({\hat\tau})$.
It is worth noting that the scale-parameters selection process via the interchangeable imputation method does not require the
model assumptions between $Y_i$ and $X_i$.
Suppose the model is set up for achieving the specific purpose mentioned above. Then  it is of interest
to incorporate the model
structure in the scale-parameters selection process. Accordingly, we propose employing a cross-validation approach to select the scale parameters. To this end,
calculate $\hat\beta_{(i)}(\tau)$
after removing the $i$-th subject from the data and compute $\hat Y_i{(\tau)}=\hat X_i^{\top}{(\tau)}\hat\beta_{(i)}(\tau)$.
Then, select $\tilde\tau$
by minimizing the squared errors of predictions across all subjects $\sum_{i=1}^n (\hat Y_i{(\tau)}-Y_i)^2$
with a grid-search over the pre-specified bounded region.

\csection{NUMERICAL STUDIES}

\csubsection{Simulation Studies}

To assess the finite sample performance of the proposed imputation method, we consider the following simulation studies.
The data are generated from a linear regression model,
$Y_i=X_i^\top\beta+\epsilon_i=X_{ia}^\top\beta_a+X_{ib}^\top\beta_b+\epsilon_i$, for $i=1,\cdots, n$,
where $\beta=(\beta_a^\top, \beta_b^\top)$.
Specifically, the discrete covariates, $X_{ia}\in\mR^2$,
are
iid simulated from a binomial distribution with probability 0.5, while
the continuous covariate, $X_{ib}\in\mR^{p-2}$ with $p=10$,
are iid generated from a
multivariate normal distribution with mean 0 and covariance matrix $\Sigma_X=(\sigma_{x,j_1j_2})$,
where $\sigma_{x,j_1j_2}=1$ for $j_1=j_2$, and $\sigma_{x,j_1j_2}=\rho$ otherwise,
and  $0<\rho<1$.
In addition, $\beta_a\in\mR^2$, $\beta_b\in\mR^{p-2}$, and $\beta_j=1$ for
$1\leq j\leq p$.
The random errors $\epsilon_i$ are iid simulated from a normal distribution with mean 0 and variance
$\sigma^2$, where $\sigma^2$ is chosen to control the coefficient of determination
$R^2=\mbox{var}(X_i^\top\beta)/\mbox{var}(Y_i)$.
We consider data
that are block-wise missing. This is a modification of the data type in Lin et al. (2020)
with finite missing patterns, which reflects our real data structure. As mentioned in
Section 2.1, we employ the Gaussian kernel of $K(\cdot)$ to construct the weight matrix defined in (2.1) throughout this entire section.

In this simulation setting, we consider  7
missing patterns: $\mathcal{D}_1=\{1,2, 3, 10\}$, $\mathcal{D}_2=\{4, 5, 6, 10\}$, $\mathcal{D}_3=\{7, 8, 9, 10\}$,  $\mathcal{D}_4=\{1,2, 3, 4, 5, 6, 10\}$, $\mathcal{D}_5=\{1,2, 3, 7, 8, 9, 10\}$, $\mathcal{D}_6=\{4, 5, 6, 7, 8, 9, 10\}$,
$\mathcal{D}_7=\{1,2, 3, 4, 5, 6, 7, 8, 9, 10\}$, where the notation of missing pattern $\mathcal{D}_k$ is defined in Appendix A for $1\leq k\leq 7$.
Accordingly, if the $i$-th sample has missing observations, its missing pattern should be one of the
aforementioned seven patterns. For example, if the $i$-th sample is missing with pattern $\mathcal{D}_1$, then $X_{ij}$ with $3<j<10$ are all missing in this sample.
Furthermore, $X_{ip}$ is observed so that it can be used to generate the ``missing at random'' data.
Such missing  patterns are similar to that of our real example.
Then we consider the three commonly used missing mechanisms:
missing completely at random (MCAR), missing at random (MAR), and missing not at random (MNAR).
Under MCAR, the missing observations in each sample are randomly generated with equal probability.
Under MAR, however, the probability of the missing observation in the $i$-th sample is
related to the observed covariate $X_{ip}$ and it is
$\{1+\mbox{exp}(-0.5X_{ip}-\alpha_1)\}^{-1}$, where $\alpha_1$ is chosen to control the
missing rate being approximately 50\%.
On the other hand, under MNAR,  the probability of the missing observation given $Y_i$ in the $i$-th sample  is related to the random error
$\epsilon_i$, which is $\{1+\mbox{exp}(-\epsilon_i-\alpha_2)\}^{-1}$ and $\alpha_2$ is selected to control the
missing rate at approximately 50\%. Lastly, the missing patterns $\mathcal{D}_j$ ($j=1, \cdots, 7$) are
randomly assigned to those missing samples.
To assess the prediction accuracy, we further randomly split the entire sample into two parts, i.e., training sample (70\%) and testing sample (30\%).

To assess the performance of SSI, we conduct 1,000 realizations.
For each realization
$k=1, \cdots, 1,000$, let
$\hat \mX^{(k)}$, $\hat\beta^{(k)}$, and $\hat Y^{(k)}_{\mbox{test}}$ be its corresponding imputed covariates,  estimate of the parameter vector, and predictions on testing samples, respectively.
In addition, let $\mX$, $\beta$, and $Y_{\mbox{test}}$ be the true covariates, true parameter vector,
and true responses from testing samples, respectively.
We next employ $\mbox{IA}=1000^{-1}\sum_k\|\hat\mX^{(k)}-\mX\|$, $\mbox{EA}=1000^{-1}\sum_k\|\hat\beta^{(k)}-\beta\|$,
and $\mbox{PA}=1000^{-1}\sum_k\|\hat Y^{(k)}_{\mbox{test}}-Y_{\mbox{test}}\|$
to correspondingly assess
the accuracy of imputations, regression estimates, and predictions.
For the sake of comparison, we also include
the commonly used methods, KNN, RF, MI, RI and FIME in our study. Here, KNN is implemented with the R package ``caret", RF
is implemented with the R package ``missForest",
RI and MI are implemented with the R package ``mice", and FIME is implemented with
the R package ``lavvan".
It is worth noting that KNN is not computable in some realizations due to
high missing rate data. Hence, we only report  computable results for KNN.
In addition, ``mice" is iterated for 50 times by burning-in the first 49 iterations to assure its stability. The rest of other methods are all implemented  by  using their corresponding default settings in
R packages.
Moreover, we use the interchangeable method and the cross-validation procedure mentioned in Section 2.3 to calculate the scale parameters in SSI
  by using a simple grid searching method with
$\tau$ ranging from 0 to 2, and we denote them  SSI$_1$ and SSI$_2$, respectively.
 It is worth noting that SSI is a weighted average of the observed values and their associated weights are evaluated by using both observed and unobserved
 information. Hence, SSI is more informative than that of the mean imputation method, which is not included in simulation studies.

In the above setting, we use  five components,
$R^2$, $\rho$, $n$, missing mechanism, and accuracy measure, to study the performance of six imputation methods.
As a result, there are various combinations via these five components. To save space,  we only present two tables and show the averaged performance.
Table 3 presents the results of IA, EA and PA under the
three coefficients of determination  ($R^2=0.3$, 0.6 and 0.9) and three missing mechanisms (MCAR, MAR, and MNAR), and they are obtained by averaging over
the three correlations ($\rho=0.25$, 0.5 and 0.75) and three sample sizes ($n=500$, 1,000 and 2,000).
Note that FIME is only applicable for parameter estimation; hence, we display its IA and PA as ``-".
In imputation,  SSI$_{1}$ performs the best, and
 SSI$_{1}$ is generally superior to SSI$_{2}$ in estimation when $R_2=0.3$ and $R_2=0.6$.
In addition, both of them  outperform the other four methods in both imputation and estimation.
As for prediction,  SSI$_{2}$ performs the best, and it is  superior to  SSI$_{1}$.  This finding is not surprising since SSI$_{1}$
is mainly designed for imputation, while SSI$_{2}$ focuses on prediction.
In sum, SSI$_{1}$ performs the best in imputation and well  in estimation, while SSI$_{2}$ outperforms other methods in prediction.
Finally,  both SSI$_{1}$ and  SSI$_{2}$ improve as $R^2$ increases. This finding is sensible since a stronger signal can lead to better imputation, estimation and prediction.

Next, Table 4 reports the results of three evaluation measures (IA, EA and PA) under
the three correlation structures ($\rho=0.25$, 0.5 and 0.75) and three missing mechanisms (MCAR, MAR and MNAR), and they are obtained by averaging over the three coefficients of
determination ($R^2=0.3$, 0.6 and 0.9)
and three sample sizes ($n=500$, 1,000 and 2,000). Analogous to the results in Table 3, both SSI$_{1}$ and  SSI$_{2}$
outperform the other methods in imputation and estimation. Although SSI$_{1}$ is slightly better than SSI$_{2}$ in imputation, SSI$_{2}$ performs the best in prediction.
Note that both SSI$_{1}$ and  SSI$_{2}$ improve in imputation and prediction as
$\rho$ becomes larger, but their performance gets worse in estimation. This finding indicates that multicollinearity can affect estimation rather than imputation and prediction, which
is expected.
Based on Tables 3 and 4, we conclude that SSI$_{1}$ and  SSI$_{2}$ are generally superior to the other four methods in imputation, estimation and prediction.
It is also worth noting that
SSI$_{1}$ and SSI$_{2}$ have low variability in the imputation bias.
This may be due to  SSI using the information from  observed responses and covariates as well as the information from covariates with missing values.
In sum, SSI not only reduces the bias of imputation, estimation and prediction, but also yields lower variability in the imputation bias.

Finally, Figure 1 depicts
the results of IA and PA for the three sample sizes ($n=500$, 1,000 and 2,000) in Panels A and B, respectively, under MNAR. The IA and PA are obtained from SSI$_{1}$ and SSI$_{2}$ by averaging over
the three correlations ($\rho=0.25$, 0.5 and 0.75) and
three coefficients of determination  ($R^2=0.3$, 0.6 and 0.9).
The results show that SSI$_{1}$ performs slightly  better (weaker) than SSI$_{2}$ in imputation (prediction) as demonstrated in Tables 3 and 4. Additionally,  they are comparable in estimation, but the EA plot is not presented here. These results also indicate that the performance
of SSI$_{1}$ and SSI$_{2}$ improves as the sample size increases,  which supports our theoretical findings.
However, this improvement is limited. Our results show that SSI is steady even with $n=500$.
Under MCAR and MAR, we obtain similar results, which are omitted.

Upon the anonymous reviewers' suggestions, we assess the robustness of the proposed  method against the non-normality of  covariates, the selection of tuning parameters, and different missing mechanisms. To this end, we consider four additional
simulation settings:
(I) the covariate $X_i$s being generated from a multivariate exponential distribution;
(II) the selection
of the tuning parameter $\tau$;
(III) the ``non-ignorable missing data'' setting modified from Kim and Yu (2011);
(IV) the missing data with no pattern. The detailed simulation settings and results are given in Appendix C of the supplementary material.
The results show similar patterns to those in Tables 3 and 4, which demonstrate the robustness of
our method.

\csubsection{Real Data Analysis}

To illustrate the usefulness of SSI, we consider a real example for risk management in online consumer finance, which was mentioned in the Introduction.
The online consumer finance industry  in China is growing rapidly in recent years and its market size is larger than 10 trillion.
Unlike the traditional personal
loan application process at commercial banks, verification of repayment ability is not feasible for
online lenders since it would time consuming for applicants to prepare and submit documents, and also
technically complicated for the credit officers to validate them in a short time.
Hence, how to measure customers' repayment ability
via their personal attributes becomes critical.

Table 1 reports the data set collected from the five different resources with six missing patterns as well as the sample size and missing rate of each individual missing pattern.
There is a total of $n=$2,390 applicants and thirteen covariates, while the response variable is the log-income. There are eleven continuous covariates and two discrete covariates. One of the discrete covariates has four class labels and the other has three class labels.
Detailed descriptions of the
thirteen covariates are presented in Table 2. Accordingly, this is a typical block-wise missing data with high missing rates; the total missing rate is 49.4\% and only 4.7\% of applicants have complete information.
To assess the accuracy of predictions,
 we randomly split the entire data set into a training sample (70\%) and
testing samples (30\%). This procedure is repeated 100 times. In the $m$-th splitting,
we calculate the mean of the
prediction error in the testing sample based on the model estimated in the training sample data, and we denote it
$\mu^{(m)}$. Then, we compute the mean and standard error of $\mu^{(m)}$ as $\mbox{MEAN}=100^{-1}\sum_{m=1}^{100} \mu^{(m)}$ and
$\mbox{SD}=\big\{100^{-1}\sum_{m=1}^{100} (\mu^{(m)}-\mbox{MEAN})^2\big\}^{1/2}$.
The scale-parameter vector $\lambda$ of SSI is
estimated via the interchangeable algorithm and the cross-validation methods, respectively, from the training sample. As in simulation studies, we denote them SSI$_{1}$ and  SSI$_{2}$.

In addition to SSI$_{1}$ and  SSI$_{2}$, we also consider the other four methods, KNN, RF, MI and RI.
Table 5 reports the MEAN and SD of all five methods. It shows that SSI$_2$ has the smallest MEAN and  SD as found in the simulation studies, while it is only slightly better than SSI$_1$.
For the sake of comparison,  we use SSI$_2$ as a baseline to calculate the percentage of reduction (PR) in MEAN and SD by comparing the other
four methods to it. Table 5 demonstrates that SSI$_2$
 yields more than 8\% reductions in MEAN and SD, respectively, over each of other methods  except for SSI$_1$.
Based on the above findings, we conclude that SSI  produces reliable predictions. Accordingly, financial institutions can multiply the predicted income  by
a pre-specified constant to determine the optimal  loan credit of each applicant. This empirical example  suggests
that SSI could play an important role in the  online consumer finance industry.

\csection{CONCLUDING REMARKS}

This paper proposes a semi-supervised imputation (SSI) method
for data that have fully observed responses and covariates with
high missing rates. The SSI is a model free method, which does not require any model assumptions.
In addition, SSI is a computationally efficient method because it has a closed form when covariates are continuous and it can be calculated  iteratively for discrete covariates.
Moreover, we show the missing-averaged consistency of SSI, and we introduce an interchange algorithm to estimate the scale parameters of SSI. After filling in missing values, we establish the relationship
between the responses and covariates via a linear regression model and introduce a cross-validation method to estimate scale parameters of SSI.  Both
simulation studies and a real application in online consumer finance
demonstrate that SSI  performs well.

It is worth noting that SSI is applicable when the response $\mY$ is partially observed. In this case,
the weight matrix $\bar A=(\bar a_{i_1i_2})\in\mR^{n\times n}$
can be defined as $\bar a_{i_1i_2}=K_{h_1}(Y_{i_1}-Y_{i_2})\prod_{k\in D_{i_1}\cap D_{i_2}}K_{h_2}(X_{i_1k}-X_{i_2k})$ if $Y_{i_1}$ and
$Y_{i_2}$ are both observed, otherwise $\bar a_{i_1i_2}=\prod_{k\in D_{i_1}\cap D_{i_2}}K_{h_2}(X_{i_1k}-X_{i_2k})$ if either
$Y_{i_1}$ or $Y_{i_2}$ is missing.
To impute missing observations in any given covariate,  SSI does not utilize other covariates' information. Hence,
we can follow an anonymous reviewer's suggestion to adapt the MI approach and
perform the SSI imputation sequentially. Specifically, we first impute $\mX_1$, and then impute
$\mX_2$ by taking into account the imputed $\mX_1$.
Repeat this procedure until $\mX_p$ has been imputed. Subsequently, we restart a new iterative procedure, and begin to
impute $\mX_1$  by taking into account the imputed $\mX_2, \cdots, \mX_p$ obtained from the previous iterative process. Then impute $\mX_2$  by taking into account the imputed $\mX_1, \mX_3,\cdots, \mX_p$. Repeat this new iterative procedure $m$ times, where $m$ is a pre-specified number of
iterations. We name this  procedure Sequentially Semi-Supervised Imputation (SSSI), and our simulation results indicate that SSSI
is slightly better than SSI; see Section D in the supplementary material.

To conclude this article, we identify five possible research avenues for
future study.  First, establish a theoretical framework to investigate the properties of SSSI.
Second, extend SSI to discrete responses so that it can be used for classification.
Third, study the efficiency and theoretical properties of imputation, estimation and prediction.
Fourth, select the optimal kernel function and its parameters automatically for the weight matrix. Lastly,
incorporate the relationship between the covariates (including the response variable) into the kernel function to construct weight matrix and improve efficiency.
We believe these efforts can broaden the usefulness of
our proposed SSI method.

\scsection{REFERENCES}

\begin{description}
\newcommand{\enquote}[1]{``#1''}
\expandafter\ifx\csname natexlab\endcsname\relax\def\natexlab#1{#1}\fi


 \bibitem[{Allison(2012)}]{Allison:2012}
Allison, P.~D. (2012).
\enquote{Handling missing data by maximum likelihood,}
 \textit{SAS Global Forum Proceedings}, 1--21.

 \bibitem[{Belkin et al.(2006)}]{Belkin:2006}
Belkin, M., Niyogi, P. and Sindhwani, V. (2006).
\enquote{Manifold regularization: A geometric framework for learning from
labeled and unlabeled examples,}
 \textit{Journal of Machine Learning Research}, 7, 2399--2434.

\bibitem[{Bollinger and Hirsch(2013)}]{Bollinger:Hirsch:2013}
Bollinger, C. and Hirsch, B. (2013).
\enquote{Is earnings nonresponse ignorable,}
 \textit{The Review of Economics and Statistics}, 95, 407--416.

  \bibitem[{Chen and Haziza(2017)}]{Chen:Haziza:2017}
Chen, S. and Haziza, D. (2017).
\enquote{Multiply robust imputation procedures for the treatment of item nonresponse
in surveys,}
 \textit{Biometrika}, 104, 439--453.

\bibitem[{Chen et al.(2015)}]{Chen:Wan:Zhou:2015}
Chen, X., Wan, A. and Zhou, Y. (2015).
\enquote{Efficient quantile regression analysis with missing
observations,}
 \textit{Journal of the American Statistical Association,}, 110, 723--741.

\bibitem[{Chapelle, Scholkopf and Zien (2010)}]{Chapelle:Scholkopf:Zien:2010}
Chapelle, O., Scholkopf, B. and Zien, A. (2010).
\textit{Semi-Supervised Learning,} Cambridge: The MIT Press.

  \bibitem[{Collins et al.(2001)}]{Collins:2001}
Collins, L. M., Schafer, J. L. and Kam, C. M. (2001).
\enquote{A comparison of inclusive and restrictive strategies
in modern missing data procedures,}
 \textit{Psychological Methods,} 23, 330--351.

\bibitem[{Du and Zhao(2017)}]{Du:Zhao:2017}
Du, C. and Zhao, Y. (2017).
\enquote{On consistency of graph-based semi-supervised learning,}
 \textit{arXiv: Machine Learning}.

 \bibitem[{Fan et al.(1997)}]{Fan:1997}
Fan, J., Gijbels, I. and King, M. (1997).
\enquote{Local likelihood and local partial likelihood in
hazard regression,}
 \textit{Annals of Statistics}, 25, 1661--1690.


\bibitem[{Fang et al.(2019)}]{Fang:2019}
Fang, F., Lan, W., Tong, J. and Shao, J. (2019).
\enquote{Model averaging
for prediction with fragmentary data,}
 \textit{Journal of Business \& Economic Statistics}, 37, 517--527.

 \bibitem[{Han and Wang(2013)}]{Han:Wang:2013}
Han, P. and Wang, L. (2013).
\enquote{Estimation with missing data: beyond double robustness,}
 \textit{Biometrika}, 100, 417--430.

\bibitem[{Kim and Yu(2011)}]{Kim:Yu:2011}
Kim, J. and Yu, L. (2011).
\enquote{A semiparametric estimation of mean functionals with nonignorable missing data,}
 \textit{Journal of the American Statistical Association}, 106, 157--165.



\bibitem[{Lafferty and Wasserman(2007)}]{Lafferty:Wasserman:2007}
Lafferty, J. and Wasserman, L. (2007).
\enquote{Statistical analysis of semi-supervised regression,}
 \textit{Advances in Neural Information Processing Systems}, 801--808.

\bibitem[{Litter(1992)}]{Litter:1992}
Little, R.~J.~A. (1992).
\enquote{Regression with missing $X$': A review,}
 \textit{Journal of the American Statistical Association}, 87, 1227--1237.

\bibitem[{Little and Rubin(2002)}]{Litter:Rubin:2002}
Little, R.~J.~A. and Rubin, D.~B. (2002).
\textit{Statistical analysis with missing data (Second Edition),} New York: Wiley.

 \bibitem[{Lin et al.(2020)}]{Lin:Liu:Lan:2020}
Lin, H., Liu, W. and Lan, W. (2020).
\enquote{Regression analysis with individual-specific
patterns of missing covariates,}
 \textit{Journal of Business \& Economic Statistics}, In Press.


  \bibitem[{Lipsitz et al.(1999)}]{Lipsitz:1999}
 Lipsitz, S. R., Ibrahim, J. G. and Zhao, L. P. (1999).
\enquote{A weighted estimating equation for missing covariate
data with properties similar to maximum likelihood,}
 \textit{Journal of the American Statistical Association}, 94, 1147--1160.


  \bibitem[{Myrtveit et al.(2001)}]{Myrtveit:2001}
Myrtveit, I., Stensrud, E. and Olsson, U. H. (2001).
\enquote{Analyzing data sets with missing data: an empirical evaluation of imputation methods and likelihood-based methods,}
\textit{IEEE Transactions on Software Engineering}, 27, 999--1013.

  \bibitem[{Nulty(2008)}]{Nulty:2008}
 Nulty, D.~D. (2008).
\enquote{The adequacy of response rates to online and paper surveys:
what can be done?}
 \textit{Assessment \& Evaluation in Higher Education}, 33, 301--314.

\bibitem[{Rajagopalan and Lall(1995)}]{Rajagopalan:1995}
Rajagopalan, B. and Lall, U. (1995).
\enquote{A kernel estimator for discrete distributions,}
 \textit{Journal of Nonparametric Statistics}, 4, 409--426.

  \bibitem[{Robins et al.(1994)}]{Robins:Rotnitzky:Zhao:1994}
Robins, J. M., Rotnitzky, A. and Zhao, L. P. (1994).
\enquote{Estimation of regression coefficients when some
regressors are not always observed,}
 \textit{Journal of the American Statistical Association}, 89, 846--866.

\bibitem[{Shao and Wang(2002)}]{Shao:Wang:2002}
Shao, J. and Wang, H. (2002).
\enquote{Sample correlation coefficients based on survey data under regression imputation,}
 \textit{Journal of American Statistical Association}, 97, 544-552.

\bibitem[{Stekhoven and B\"uhlmann(2012)}]{Stekhoven:2012}
Stekhoven, D. and B\"uhlmann, P. (2012).
\enquote{MissForest non-parametric missing value imputation for mixed-type data,}
 \textit{Bioinformatics}, 28, 112--118.

 \bibitem[{Troyanskaya et al.(2001)}]{Troyanskaya:2001}
Troyanskaya, O., Cantor, M., Sherlock, G., Brown, P., Hastie, T., Tibshirani, R., Botstein, D. and Altman, R. (2001).
\enquote{Missing value estimation methods for DNA microarrays,}
 \textit{Bioinformatics}, 17, 520--525.

\bibitem[{Xiang et al.(2014)}]{Xiang:Yuan:Fan:Wang:Thompson:Ye:2014}
Xiang, S., Yuan, L., Fan, W., Wang, Y., Thompson, P. M. and Ye, J.  (2014).
\enquote{Bi-level multi-
source learning for heterogeneous block-wise missing data,}
 \textit{Neuroimage}, 102, 192--206.

 \bibitem[{Yuan et al.(2012)}]{Yuan:Wang:Thompson:Narayan:Ye:2012}
Yuan, L., Wang, Y., Thompson, P. M.,  Narayan, V. A. and Ye, J.  (2012).
\enquote{Multi-source feature learning for joint analysis of incomplete multiple heterogeneous
neuroimaging data,}
 \textit{Neuroimage}, 61, 622--632.


\bibitem[{Zhang and Li(2006)}]{Zhang:Li:2006}
Zhang, X. and Li, W.~S. (2006).
\enquote{Hyperparameter learning for graph based semi-supervised
learning algorithms,}
 \textit{Advances in
Neural Information Processing Systems}, 19, 1595--1592.

\bibitem[{Zhang(2016)}]{Zhang:2016}
Zhang, Z. (2016).
\enquote{Missing data imputation: focusing on single imputation,}
 \textit{Annals of Translational Medicine}, 4, 1--8.

%

  \bibitem[{Zhou et al.(2010)}]{Zhou:Litter:Kalbeisch:2010}
Zhou, Y., Litter, R.~J.~A. and Kalbeisch, J.~D. (2010).
\enquote{Block-conditional missing at random
models for missing data,}
 \textit{Statistical Science}, 25, 517--532.

  \bibitem[{Zhu et al.(2003)}]{Zhu::Ghahramani:Lafferty:2003}
Zhu, X., Ghahramani, Z. and Lafferty, J. (2003).
\enquote{Semi-supervised learning using Gaussian
fields and Harmonic functions,}
 \textit{International Conference on Machine Learning}, 118, 912--919.

\bibitem[{Zhu and Goldberg (2009)}]{Zhu:Goldberg:2009}
Zhu, X. and Goldberg, A.~B. (2009).
\textit{Introduction to Semi-Supervised Learning,} San Rafael: Morgan \& Claypool.

 \bibitem[{Zou et al.(2017)}]{Zou:Lan:Wang:Tsai:2017}
Zou, T., Lan, W., Wang, H. and Tsai., C. L. (2017).
\enquote{Covariance regression analysis,}
 \textit{Journal of the American Statistical Association}, 112, 266--281.

\end{description}

\newpage

\begin{table}
\begin{center}
\begin{small}
\caption{Sample sizes and missing rates of the six missing patterns for the empirical data,
where ``1" represents the data being available in the corresponding source, and ``0" otherwise.}
\label{table:t1}
\begin{tabular}{c|ccccc|cc}
\hline \hline Pattern & Credit & e-shopping record & Cell phone &
Credit bureau
& Fraud &Sample & Missing \\
 &card& in Taobao& usage
record&
information& record& size & rate (\%)\\
\hline
1&1&1&1&1&1&113& 0.0\\
2&1&1&1&1&0&29 & 15.4\\
3&1&1&0&1&1&231 & 15.4\\
4&1&1&1&0&0&220 & 30.8\\
5&1&1&0&0&0&1,161 & 46.2\\
6&0&1&0&0&0&636 & 84.6\\
\hline
  & & & & &Total&2,390 & 49.4\\
\hline
\end{tabular}
\end{small}
\end{center}
\end{table}

\begin{landscape}
\begin{table}
\begin{center}
\begin{small}
\caption{Detailed descriptions of the 13 covariates including data type, data source and variable definition. Note that
 the variables ``credit-use-number" and ``credit-use-amount" are discrete, and they have four and three class labels, respectively.}
\label{table:t1}
\begin{tabular}{cccc}
\hline \hline
Covariate & Data Type &Data Source & Definition\\
\hline
credit-min  & Continuous & Credit card & the minimum credit limit among all credit cards \\
credit-max  & Continuous & Credit card & the maximum credit limit among all credit cards \\
credit-union & Continuous & Credit card & the total amount of credit limit of credit cards \\
credit-use-number & Discrete (four class labels) & Credit card & average transaction number of credit cards \\
credit-use-amount & Discrete (three class labels)& Credit card & average transaction amount of credit cards \\
shopping-record-amount & Continuous & E-shopping record & average transaction amount of e-shopping \\
shopping-record-number & Continuous & E-shopping record & average transaction number of e-shopping \\
phone-use-plansum & Continuous & Cell-phone usage & total mobile package fee over the last 6 months\\
phone-use-totalsum & Continuous & Cell-phone usage & total mobile cost over the last 6 months \\
payment-amount & Continuous & Credit bureau & average payment for commercial banks \\
card-number & Continuous & Credit bureau & number of active bank cards \\
approval-number & Continuous & Fraud record & approval number for non-commercial banks \\
fraud-score & Continuous & Fraud record & fraud score evaluated by credit company \\
\hline
\end{tabular}
\end{small}
\end{center}
\end{table}
\end{landscape}

\begin{landscape}
\begin{table}[htbp!]
\begin{center}
\caption
{The simulation results of IA, EA and PA under the
three coefficients of determination ($R^2$=0.3, 0.6 and 0.9) and three missing mechanisms (MCAR, MAR, and MNAR), and they are obtained by averaging over three correlations ($\rho=0.25$, 0.5 and 0.75) and three sample sizes ($n=500$, 1,000 and 2,000). The value in parentheses is the averaged standard error.
Note that FIME is only applicable for parameter estimation, hence, we display its IA and PA as ``-".}
\label{tb:4}
\vspace{0.28 cm}
\begin{tabular}{c|c|ccc|ccc|ccc}
\hline
&  &  \multicolumn{3}{|c}{MCAR} & \multicolumn{3}{|c}{MAR} & \multicolumn{3}{|c}{MNAR} \\
$R^2$ & Methods & IA & EA  &  PA & IA & EA  &  PA & IA & EA  &  PA\\
\hline
0.3 & KNN  & 0.79(0.197) & 0.76(0.061) & 0.83(0.098) & 1.02(0.216) & 0.74(0.062) & 0.97(0.124) & 0.79(0.219) & 0.76(0.064) & 0.97(0.094) \\
    & RF   & 0.74(0.228) & 0.78(0.027) & 0.78(0.081) & 0.75(0.234) & 0.78(0.027) & 0.79(0.085) & 0.74(0.233) & 0.78(0.027) & 0.78(0.083) \\
    & MI   & 1.15(0.347) & 0.79(0.024) & 0.80(0.085) & 1.15(0.347) & 0.79(0.025)  & 0.80(0.088) & 1.15(0.351) & 0.79(0.023) & 0.79(0.088) \\
    & RI   & 0.86(0.262) & 0.81(0.141) & 0.97(0.399) & 0.86(0.237) & 1.28(0.959) & 2.12(1.962) & 0.82(0.266) & 0.94(0.498) & 1.39(1.347) \\
    & FIME  & -           & 0.76(0.088) & -            & -            & 0.76(0.048) & -            & -            & 0.76(0.042)  & -            \\
    & SSI$_1$ & 0.68(0.157) & 0.70(0.069) & 0.83(0.083) & 0.68(0.158)  & 0.70(0.075) & 0.83(0.087)  & 0.68(0.157) & 0.70(0.065) & 0.81(0.087) \\
    & SSI$_2$ & 0.70(0.149) & 0.72(0.046)  & 0.77(0.079) & 0.70(0.149) & 0.72(0.051) & 0.78(0.083) & 0.73(0.148) & 0.72(0.054) & 0.77(0.081)  \\
    \hline
0.6 & KNN  & 0.79(0.192) & 0.70(0.077) & 0.73(0.096) & 0.86(0.197) & 0.70(0.069) & 0.73(0.094) & 0.79(0.213) & 0.71(0.068) & 0.80(0.091) \\
    & RF   & 0.74(0.229) & 0.73(0.027) & 0.54(0.088) & 0.75(0.234) & 0.73(0.028) & 0.54(0.090) & 0.74(0.235) & 0.73(0.028) & 0.54(0.093) \\
    & MI   & 1.15(0.351) & 0.74(0.022) & 0.57(0.094) & 1.15(0.348) & 0.74(0.022)  & 0.57(0.092) & 1.15(0.349) & 0.74(0.021) & 0.57(0.096) \\
    & RI   & 0.86(0.260)  & 0.73(0.093) & 0.66(0.252)  & 0.88(0.255) & 0.81(0.341) & 0.89(0.894) & 0.80(0.234) & 1.13(0.723) & 1.79(2.039) \\
    & FIME  & -           & 0.69(0.045) & -            & -            & 0.69(0.048)  & -            & -            & 0.69(0.047) & -            \\
    & SSI$_1$ & 0.68(0.158) & 0.65(0.053) & 0.54(0.072) & 0.68(0.158) & 0.65(0.055) & 0.54(0.072) & 0.68(0.159) & 0.65(0.055) & 0.54(0.074) \\
    & SSI$_2$ & 0.68(0.157) & 0.66(0.053) & 0.52(0.075) & 0.68(0.157) & 0.66(0.051) & 0.52(0.072) & 0.69(0.161) & 0.65(0.058) & 0.52(0.076) \\
    \hline
0.9 & KNN  & 0.80(0.192) & 0.66(0.062) & 0.48(0.125) & 0.81(0.197) & 0.76(0.072) & 0.48(0.119) & -            & -            & -            \\
    & RF   & 0.74(0.229) & 0.69(0.028) & 0.29(0.107) & 0.75(0.234) & 0.69(0.027) & 0.31(0.111) & 0.75(0.234) & 0.69(0.027) & 0.31(0.121) \\
    & MI   & 1.15(0.349) & 0.70(0.021) & 0.34(0.121)  & 1.15(0.352)  & 0.70(0.021) & 0.34(0.119) & 1.15(0.345) & 0.71(0.061) & 0.35(0.122)  \\
    & RI   & 0.86(0.261) & 0.67(0.049) & 0.36(0.168) & 0.85(0.264) & 0.66(0.045) & 0.34(0.142) & 0.75(0.222) & 1.92(1.709) & 2.63(2.486) \\
    & FIME  & -            & 0.64(0.044)  & -            & -            & 0.64(0.043) & -            & -            & 0.64(0.045)  & -            \\
    & SSI$_1$ & 0.67(0.162)  & 0.63(0.042) & 0.27(0.083) & 0.67(0.162) & 0.63(0.041) & 0.27(0.083) & 0.67(0.161) & 0.63(0.042) & 0.28(0.087) \\
    & SSI$_2$ & 0.68(0.155) & 0.62(0.053) & 0.26(0.079) & 0.68(0.161)  & 0.62(0.053) & 0.27(0.083) & 0.68(0.155) & 0.61(0.057) & 0.27(0.083) \\
\hline
\end{tabular}
\end{center}
\end{table}
\end{landscape}

\begin{landscape}
\begin{table}[htbp!]
\begin{center}
\caption
{The simulation results of IA, EA and PA under the three correlations ($\rho=0.25$, 0.5 and 0.75) and three missing mechanisms (MCAR, MAR, and MNAR), and they are obtained by averaging over the
three coefficients of determination ($R^2$=0.3, 0.6 and 0.9) and three sample sizes ($n=500$, 1,000 and 2,000). The value in parentheses is the averaged standard error.
Note that FIME is only applicable for parameter estimation; hence, we display its IA and PA as ``-".}
\label{tb:4}
\vspace{0.28 cm}
\begin{tabular}{c|c|ccc|ccc|ccc}
\hline
 &  &  \multicolumn{3}{|c}{MCAR} & \multicolumn{3}{|c}{MAR} & \multicolumn{3}{|c}{MNAR} \\
$\rho$ & Methods & IA & EA  &  PA & IA & EA  &  PA & IA & EA  &  PA\\
\hline
0.25 & KNN  & 1.00(0.066) & 0.67(0.075) & 0.79(0.223) & 1.03(0.089) & 0.84(0.072) & 0.74(0.239) & 1.04(0.110)   & 0.80(0.074) & 0.85(0.208) \\
     & RF   & 1.01(0.056) & 0.71(0.048) & 0.62(0.178) & 1.02(0.060) & 0.71(0.051) & 0.64(0.178) & 1.02(0.059) & 0.71(0.048) & 0.63(0.175) \\
     & MI   & 1.54(0.095) & 0.73(0.042) & 0.67(0.163) & 1.54(0.094) & 0.73(0.045) & 0.67(0.165) & 1.54(0.094) & 0.73(0.043)  & 0.67(0.161) \\
     & RI   & 1.15(0.101) & 0.70(0.104) & 0.75(0.372) & 1.14(0.103) & 0.77(0.318) & 1.06(1.265) & 1.13(0.098) & 1.28(0.829) & 2.73(2.219) \\
     & FIME  & -           & 0.65(0.078)  & -            & -            & 0.65(0.061) & -            & -            & 0.65(0.067) & -            \\
     & SSI$_1$ & 0.85(0.038) & 0.60(0.038)   & 0.61(0.208) & 0.85(0.037)  & 0.60(0.036) & 0.61(0.211) & 0.85(0.038) & 0.60(0.039) & 0.61(0.203) \\
     & SSI$_2$ & 0.85(0.038) & 0.61(0.058) & 0.58(0.192) & 0.86(0.040)  & 0.61(0.056) & 0.59(0.194) & 0.86(0.039) & 0.60(0.061) & 0.59(0.192) \\
     \hline
0.5  & KNN  & 0.81(0.072)   & 0.76(0.065) & 0.72(0.411) & 0.88(0.142) & 0.89(0.054)  & 0.75(0.285) & 0.86(0.111)  & 0.73(0.061)  & 0.86(0.298) \\
     & RF   & 0.75(0.046) & 0.73(0.043) & 0.54(0.211) & 0.76(0.046) & 0.74(0.043) & 0.53(0.211) & 0.76(0.044) & 0.74(0.041) & 0.54(0.208) \\
     & MI   & 1.19(0.082) & 0.75(0.037) & 0.56(0.199) & 1.19(0.089) & 0.75(0.039) & 0.57(0.201) & 1.19(0.087)  & 0.75(0.036) & 0.57(0.195) \\
     & RI   & 0.88(0.095) & 0.74(0.099) & 0.65(0.369) & 0.89(0.098) & 0.86(0.486) & 1.02(1.448) & 0.88(0.085) & 1.01(0.587)  & 1.53(1.661) \\
     & FIME  & -            & 0.70(0.053) & -            & -            & 0.70(0.057) & -            & -            & 0.71(0.055) & -            \\
     & SSI$_1$ & 0.71(0.035) & 0.66(0.039) & 0.54(0.234) & 0.71(0.035) & 0.66(0.038) & 0.54(0.238) & 0.71(0.035) & 0.66(0.039) & 0.54(0.227) \\
     & SSI$_2$ & 0.72(0.037) & 0.67(0.047) & 0.51(0.216)  & 0.71(0.035) & 0.67(0.050)  & 0.51(0.217) & 0.72(0.042)  & 0.67(0.05)  & 0.51(0.215) \\
     \hline
0.75 & KNN  & 0.57(0.076) & 0.84(0.059) & 0.66(0.430)  & 0.68(0.217)  & 0.92(0.069) & 0.59(0.301) & 0.58(0.123) & 0.86(0.062) & 0.68(0.285) \\
     & RF   & 0.49(0.028) & 0.75(0.042) & 0.46(0.235) & 0.49(0.029) & 0.75(0.039) & 0.48(0.235) & 0.50(0.027) & 0.75(0.039) & 0.49(0.233) \\
     & MI   & 0.71(0.061) & 0.76(0.036)  & 0.48(0.227) & 0.72(0.063) & 0.76(0.038)  & 0.48(0.226) & 0.72(0.061) & 0.76(0.035)  & 0.48(0.219) \\
     & RI   & 0.55(0.076) & 0.78(0.135) & 0.58(0.387) & 0.57(0.087) & 0.98(0.808) & 0.91(1.27)  & 0.55(0.083) & 1.08(0.948) & 1.16(1.353) \\
     & FIME  & -            & 0.74(0.079) & -            & -            & 0.74(0.051) & -            & -            & 0.74(0.048) & -            \\
     & SSI$_1$ & 0.47(0.024) & 0.71(0.040) & 0.49(0.260)  & 0.47(0.027) & 0.72(0.051) & 0.48(0.259) & 0.47(0.025) & 0.71(0.044) & 0.48(0.245) \\
     & SSI$_2$ & 0.49(0.036)  & 0.72(0.041) & 0.46(0.235) & 0.49(0.038)  & 0.72(0.042) & 0.46(0.237)  & 0.51(0.101) & 0.72(0.047) & 0.46(0.234) \\
\hline
\end{tabular}
\end{center}
\end{table}
\end{landscape}

\begin{figure}[!h]
\caption{The plots of IA (panel A) and PA (panel B) versus the sample sizes $n=500$, 1,000, and 2,000  under MNAR.  The IA and PA are obtained from SSI$_{1}$ and SSI$_{2}$ by averaging over
 three correlations ($\rho=0.25$, 0.5 and 0.75) and
three coefficients of determination  ($R^2=0.3$, 0.6 and 0.9).}
\begin{center}
\includegraphics[angle=270,width=5.0 in]{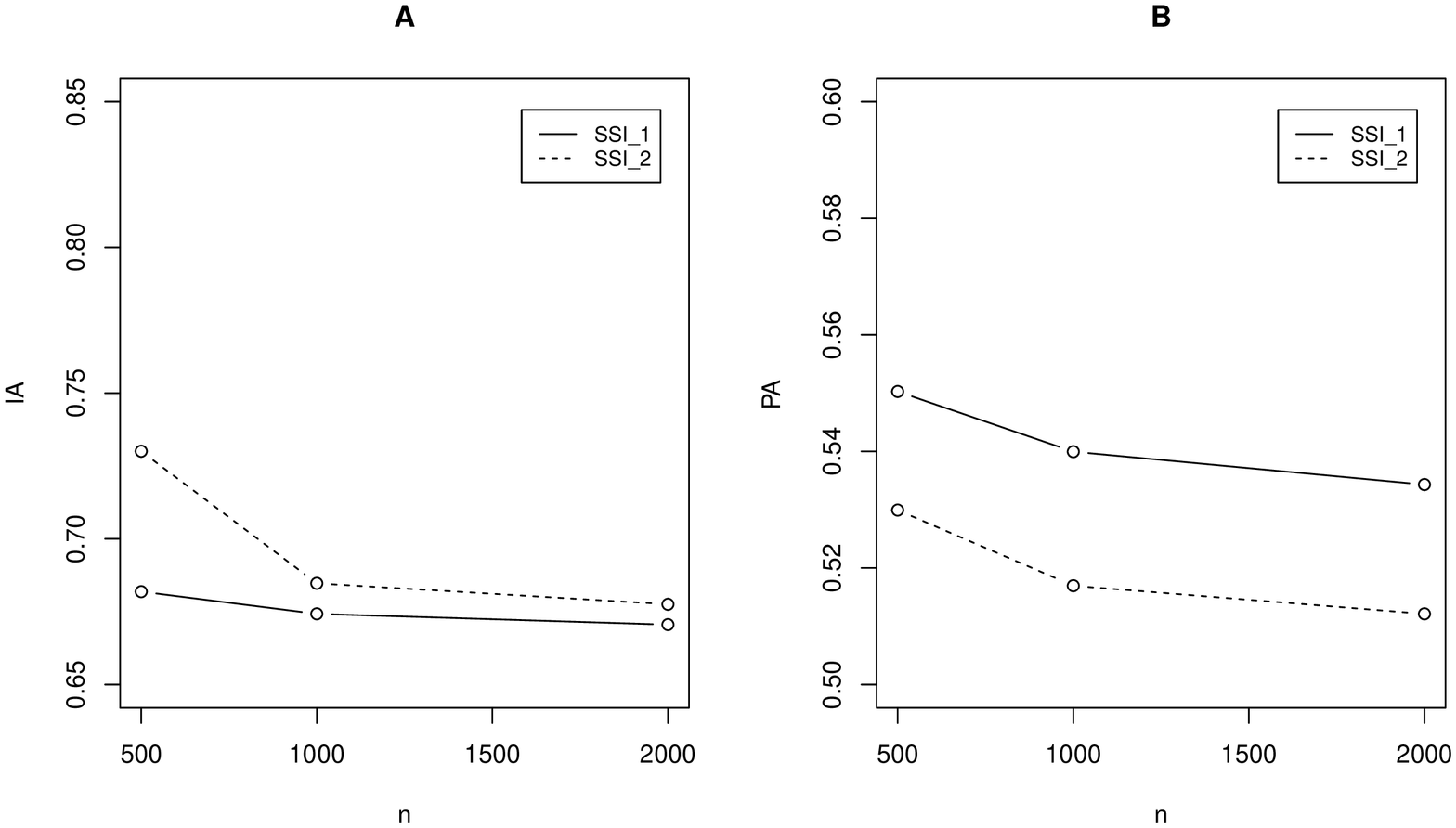}
\end{center}
\end{figure}

\begin{table}[htbp!]
\begin{center}
\caption
{The mean (MEAN) and standard error (SD) of the prediction errors obtained from KNN, MI, RI, MissForest, SSI$_1$ and SSI$_2$,
and the percentage reduction (PR) in Mean and SD that is achieved by SSI$_2$ in comparison to the given method.
}
\label{tb:4}
\vspace{0.28 cm}
\begin{tabular}{lccccccc}
\hline
 &   KNN & MI & RI & RF & SSI$_1$ & SSI$_2$ \\
\hline
Mean   &  0.311  & 0.474 &  0.360 & 0.306 & 0.282 & 0.280\\
PR (\%)  & 10.21 & 40.93 & 22.22 & 8.02 & 0.72 & 0\\
\hline
SD    &    0.053  & 0.077 &  0.070 &  0.052 & 0.049 & 0.048 \\
PR (\%) &  9.43 & 37.66 & 31.42 & 8.42 & 2.08 & 0\\
\hline
\end{tabular}
\end{center}
\end{table}

\end{document}


\begin{center}
{\bf\Large Imputations for High Missing Rate Data in Covariates via Semi-supervised Learning Approach}\\
{Supplementary Material} \\
\bigskip
\end{center}
\begin{center}
{Wei Lan, Xuerong Chen, Tao Zou and Chih-Ling Tsai}

{\it Southwestern University of Finance and Economics, The Australian National University
and University of California, Davis}
\end{center}

\begin{center}
\end{center}

This supplementary material includes four components. Appendix A
presents six technical conditions, Appendix B provides the proof of Theorem 1,
Appendix C provides additional simulations to assess the robustness of SSI against data non-normality, tuning parameter selection and different missing
mechanisms, and
Appendix D
presents simulation results for sequentially semi-supervised imputation, SSSI.
Note that the conditions are used only for the theoretical proofs, and not for the practical imputations of SSI.

\renewcommand{\theequation}{A.\arabic{equation}}
\setcounter{equation}{0}

\scsubsection{Appendix A: Technical Conditions.}

Before presenting technical conditions, we introduce some definitions and notations.
Recall that $D_i=\{j: D_{ij}=1\}$, which contains the observed covariates for subject $i$.
Thus, we define that two subjects $i_1$ and $i_2$ have the same
missing pattern if $D_{i_1}=D_{i_2}$.
Denote $d_0=1+\max_{1\leq i\leq n-1}|D_{i}\cap D_{i+1}|$, and
let $\mathcal{D}=\{\mathcal{D}_1,\cdots, \mathcal{D}_{M_0}\}$, where $\mathcal{D}_1,\cdots, \mathcal{D}_{M_0}$
are unique elements of set $D=\{D_1, \cdots, D_n\}$.
In addition, let
$\bZ$ be
the difference between two independent replicates of $(Y,X^\top)^\top$'s, i.e., $\bZ=(Y_{i_1},X_{i_1}^\top)^\top-(Y_{i_2},X_{i_2}^\top)^\top$ for any two indexes $i_1$ and
$i_2$, with the probability density function
$g_{Z}(z)$. For the sake of simplicity, we assume that $K(\cdot)$ in Equation (2.1) is a probability density function, and then we define $\mu_k=\int u^kK(u)du$.
The following conditions are used for proving Theorems 1 and 2.

\begin{itemize}

\item[(C1)]
For any $i,i^*=1,\cdots,n$, $j=1,\cdots,p$ and $m=1, \cdots, M_0$, we assume that the density function $f(Y_i,X_{ik},k\in D_i\cap D_{i^*})$,
 conditional probability $P(D_i=\mD_m|Y_i,X_{ik}, k\in D_i\cap D_{i^*})$ and conditional expectation $E(X_{ij}|Y_i,X_{ik}, k\in D_i\cap D_{i^*})$
are all bounded and have continuous $b$-th derivatives with $b\geq 2$ and $2b>2d_0+1$. Here, the conditional expectation is respect to the conditional density function
$f(X_{ij}|Y_i,X_{ik}, k\in D_i\cap D_{i^*})$.

\item[(C2)] $K_{h_l}(\cdot)$ is a kernel function of order $b$ with compact support, and it satisfies the Lipschitz condition. In addition, assume that
 $h_l\to 0$,  $nh_l^{2d_0}\to \infty$, $nh_l^{2b}\to 0$ and $nh_l^{d_0}/\log n\to \infty$, for  $l=1$ and $2$, as $n\to \infty$.

 \item[(C3)] Assume that $M_0<\infty$, where $M_0$ is the number of distinct missing patterns. In addition, there exists a constant $c_0\in(0,1)$ such that $|\mS_{1j}|/n\to c_0$ for $j=1,\cdots,p$.

\item[(C4)] For any $i\in\mS_{0j},$ assume that
  $\sum_{i^*\in\mS_{1j}} a_{ii^*}/\sum_{l=1}^n a_{il}\to_p 1$.

\item[(C5)] Assume that there exist finite positive constants $\delta_0$ and $\delta_1$ such that, for any $\Delta\leq \delta_0$, $\min_{\|z\|\leq\Delta}g_{Z}(z)=\delta_1>0$.

\item[(C6)] For any $i\in \mS_{0j}$, $i^*\in \mS_{1j}$, $j=1,\cdots,p$ and $m=1,\cdots,M_0$,
assume that  $E\{I(D_{i}=\mD_m)X_{i^*j}|Y_{i^*},X_{i^*k}, k\in D_i\cap D_{i^*}\}=P(D_i=\mD_m|Y_{i^*},X_{i^*k}, k\in D_i\cap D_{i^*})E\{X_{i^*j}|Y_{i^*},X_{i^*k}, k\in D_i\cap D_{i^*}\}$,
 where the conditional
expectation is with respect to the conditional density function $f(X_{ij}|Y_i,X_{ik}, k\in D_i\cap D_{i^*})$.

\end{itemize}

Similar assumptions to (C1)-(C2) have been used in the nonparametric literature (see, e.g., Fan et al., 1997; Chen et al., 2015).
For example, our assumption for the density function is similar to Condition (C1) of Chen et al. (2015) and Condition A (vi)-(vii) of Fan et al. (1997).
Condition (C3)
controls the number of the  missing patterns of the data,  and it holds when the data is block-wise missing (Xiang et al., 2014; Lin et al., 2019).
This condition ensures that there are enough samples
being used to impute missing covariates and thus resulting imputations can be accurate.
Condition (C4) is a technical condition to assure the summation of weights of $C_{\mS_0j}$ (the leading term of $\hat X_{\mS_{0j}j}$)
is approaching to 1.
Accordingly,
the observed subjects  are given relatively more weight than the missing subjects.
Condition (C5) is the same as that assumed in
Theorem 1 of Du and Zhao (2017),
which  controls the norm of $W_{\mS_{0j}}$ to assure the
convergence of iterative imputations.
 Condition (C6) is a
technical condition which is used
to show Theorems 1 and 2.
Note that this condition holds if the missing pattern $D_i$ is independent of $Y_i$ and $X_i$.
Hence, Condition (C6) is satisfied when the data is missing completely at random (MCAR).
In addition, Condition (C6) differs from the
missing at random (MAR) assumption $P(D_i=\mD_m|Y_i, X_i)=P(D_i=\mD_m|Y_i)$ for any $m=1, \cdots, M_0$. An example is when $D_i$ is independent of
$X_i$ but correlated with the random error $\epsilon_i$ via the linear regression model setting, $Y_i=X_i^\top\beta+\epsilon_i$. Then Condition (C6) holds,
while $P(D_i=\mD_m|Y_i, X_i)\not=P(D_i=\mD_m|Y_i)$. The latter occurs because $X_i$ is correlated with $\epsilon_i=Y_i-X_i^\top\beta$ given $Y_i$.
Accordingly, $X_i$ may be correlated with $D_i$ given $Y_i$.
In this case, Condition (C6) is missing not at random (MNAR).

\scsubsection{Appendix B: Proof of Theorem 1}

Without loss of generality, we only consider the proof with $p=2$. Accordingly, there are two covariates
$X_{i1}$ and $X_{i2}$ for each subject $i$ and three missing patterns $\mathcal{D}_1=\{1\}$, $\mathcal{D}_2=\{2\}$, and $\mathcal{D}_3=\{1,2\} $, which
$D_i\in \mathcal{D}=:\{\mathcal{D}_1,\mathcal{D}_2,\mathcal{D}_3\}$.
In addition, $d_0=2$ so that $b=3$ can meet the requirement of Condition (C1).
By Condition (C2), we then have
$nh_l^4\to \infty$, $nh_l^6\to 0$ for $l=1$ and $2$.

Define $\mathcal{W}_m=W_{\mS_{0j}}+(W_{\mS_{0j}})^{2}+\cdots+(W_{\mS_{0j}})^{m}$.
Under Conditions (C2) and (C5), we then apply similar techniques to those used in the proof of Theorem 1 of Du and Zhao (2017) to show that
$\mathcal{W}=:\lim_{m\to \infty}\mathcal{W}_m$,
\[
    (I-W_{\mS_{0j}})^{-1}=I+\lim_{m\to \infty}\mathcal{W}_m=I+\mathcal{W},\,\, and ~~  \|\mathcal{W}\|_{\max}=O_p((nh^{d_0})^{-1}),
\]
with probability approaching 1 as $n\rightarrow \infty$, where $\|A\|_{\max}=\max_{i,j}a_{ij}$ for any arbitrary matrix $A=(a_{ij})$.
For the sake of simplicity, we do not state ``with probability approaching 1'' in the rest of proof.

By (2.4), we have $\hat X_{\mS_{0j}j}=(I-W_{\mS_{0j}})^{-1}C_{\mS_{0j}}$, which implies that
\begin{eqnarray*}
    \hat X_{\mS_{0j}j}=(I-W_{\mS_{0j}})^{-1}C_{\mS_{0j}}=(I+\mathcal{W})C_{\mS_{0j}}=C_{\mS_{0j}}+\mathcal{W}C_{\mS_{0j}}.
\end{eqnarray*}
Accordingly, we  obtain that
\begin{equation}
    \hat X_{\mS_{0j}j}-X_{\mS_{0j}j}
    =(C_{\mS_{0j}}-X_{\mS_{0j}j})+\mathcal{W}C_{\mS_{0j}}
    =:J_1+J_2=(J_{1i})_{i\in\mS_{0j}}+(J_{2i})_{i\in\mS_{0j}}.
\end{equation}

We first consider $J_{1i}$ in (A.1) for
$i\in \mS_{0j}$.
By Condition (C4), we have
\begin{equation}
    J_{1i}=\frac{\sum_{i^*\in\mS_{1j}} a_{ii^*} X_{i^*j}}{\sum_{i^*=1}^n a_{ii^*}}-X_{ij}
    =\frac{\sum_{i^*\in\mS_{1j}} a_{ii^*} X_{i^*j}}{\sum_{i^*\in\mS_{1j}} a_{ii^*}}\{1+o_p(1)\}-\frac{\sum_{i^*\in\mS_{1j}} a_{ii^*}X_{ij}}{\sum_{i^*\in\mS_{1j}} a_{ii^*}}.
\end{equation}
Without loss of generality, it suffices to show that the first covariate $X_{i1}$ has missing values.
Hence, for any $i\in\mS_{01} $,
$D_i=\mD_2$.
For the sake of convenience, denote $P(D=\mathcal{D}_m|Y=Y_i,X_j=X_{ij})$ and
$E(X_{1}|Y=Y_i,X_j=X_{ij})$
as  $P(D=\mathcal{D}_m|Y_i,X_{ij})$ and  $E(X_{1}|Y_i,X_{ij})$, respectively, where $D$,
$X_1$ and $X_j$ represent their  corresponding population of $D_i$, $X_{i1}$ and $ X_{ij}$.

There are  two components, $\sum_{i^*\in\mS_{11}} a_{ii^*} X_{i^*1}$ and
$\sum_{i^*\in\mS_{11}} a_{ii^*}$, in the first term of the right hand side of  (A.2).
After algebraic simplification, the first component is
\begin{eqnarray*}
	\sum_{i^*\in\mS_{11}} a_{ii^*} X_{i^*1}&=&I(D_i=\mD_2)\sum_{i^*\in\mS_{11}} a_{ii^*} X_{i^*1}\\
	&=&I(D_i=\mD_2)\bigg[\sum_{i^*\in\mS_{11}}I(D_{i^*}=\mathcal{D}_1)K_{h_1}(Y_i-Y_{i^*})X_{i^*1}\\&&+\sum_{i^*\in\mS_{11}}I(D_{i^*}=\mathcal{D}_2)K_{h_1}(Y_i-Y_{i^*})K_{h_2}(X_{i2}-X_{i^*2})X_{i^*1}\\ &&
+\sum_{i^*\in\mS_{11}}I(D_{i^*}=\mathcal{D}_3)K_{h_1}(Y_i-Y_{i^*})K_{h_2}(X_{i2}-X_{i^*2})X_{i^*1}\bigg].
\end{eqnarray*}
Under Conditions (C1), (C2) and (C3), we then apply similar techniques to those used in the proof of  Chen et al. (2015), by letting $(Y_{i^*}-Y_i)/h_1=u$, and we obtain
\begin{eqnarray*}
	&&\sum_{i^*\in\mS_{11}}I(D_{i^*}=\mathcal{D}_1)K_{h_1}(Y_i-Y_{i^*})X_{i^*1}\\
	&=&|\mS_{11}|\bigg[E\big\{E[I(D_{i^*}=\mathcal{D}_1)K_{h_1}(Y_i-Y_{i^*})X_{i^*1}]|Y_{i^*}\big\}+O_p\big\{(\frac{\log h_1^{-1}}{nh_1^2})^{1/2}\big\}\bigg]\\
	&=&|\mS_{11}|\bigg[\int P(D=\mathcal{D}_1|Y_{i^*})E(X_{1}|Y_{i^*})K_{h_1}(Y_{i^*}-Y_i)f(Y_{i^*})dY_{i^*}+O_p\big\{(\frac{\log h_1^{-1}}{nh_1^2})^{1/2}\big\}\bigg]\\
	&=&|\mS_{11}|\bigg[\int P(D=\mathcal{D}_1|Y_i+h_1u)E(X_{1}|Y_i+h_1u)K(u)f(Y_i+h_1u)du+O_p\big\{(\frac{\log h_1^{-1}}{nh_1^2})^{1/2}\big\}\bigg]\\
	&=&|\mS_{11}|\bigg[P(D=\mathcal{D}_1|Y_i)E(X_{1}|Y_i)f(Y_i)+h^3_1\mu_3\frac{\partial [P(D=\mathcal{D}_1|Y_i)E(X_{1}|Y_i)f(Y_i)] }{\partial Y_i}\\
	&&+O_p\{h_1^4+(\frac{\log h_1^{-1}}{nh_1^2})^{1/2}\}\bigg],
\end{eqnarray*}
where the second equality is due to Condition (C6).
Analogously, we can obtain that
\begin{eqnarray*}
	&&\sum_{i^*\in\mS_{11}}I(D_{i^*}=\mathcal{D}_2)K_{h_1}(Y_i-Y_{i^*})K_{h_2}(X_{i2}-X_{i^*2})X_{i^*1}\\
	&=&|\mS_{11}|\bigg[P(D=\mathcal{D}_2|Y_i,X_{i2})E(X_{1}|Y_i,X_{i2})f(Y_i,X_{i2})\\
	&&+h^3_1\mu_3\frac{\partial^3 [P(D=\mathcal{D}_2|Y_i,X_{i2})E(X_{1}|Y_i,X_{i2})f(Y_i,X_{i2})] }{\partial Y_i^3}\\
	&&+h^3_2\mu_3\frac{\partial^3 [P(D=\mathcal{D}_2|Y_i,X_{i2})E(X_{1}|Y_i,X_{i2})f(Y_i,X_{i2})] }{\partial X_{i2}^3}\\&&+O_p\{h_1^4+(\frac{\log h_1^{-1}}{nh_1^2})^{1/2}\}+O_p\{h_2^4+(\frac{\log h_2^{-1}}{nh_2^2})^{1/2}\}\bigg],
\end{eqnarray*}
and
\begin{eqnarray*}
	&&	\sum_{i^*\in\mS_{11}}I(D_{i^*}=\mathcal{D}_3)K_{h_1}(Y_i-Y_{i^*})K_{h_2}(X_{i2}-X_{i^*2})X_{i^*1}\\
	&=&|\mS_{11}|\bigg[P(D=\mathcal{D}_3|Y_i,X_{i2})E(X_{1}|Y_i,X_{i2})f(Y_i,X_{i2})\\
	&&+h^3_1\mu_3\frac{\partial^3 [P(D=\mathcal{D}_3|Y_i,X_{i2})E(X_{1}|Y_i,X_{i2})f(Y_i,X_{i2})] }{\partial Y_i^3}\\
	&&+h^3_2\mu_3\frac{\partial^3 [P(D=\mathcal{D}_3|Y_i,X_{i2})E(X_{1}|Y_i,X_{i2})f(Y_i,X_{i2})] }{\partial X_{i2}^3}\\&&+O_p\{h_1^4+(\frac{\log h_1^{-1}}{nh_1^2})^{1/2}\}+O_p\{h_2^4+(\frac{\log h_2^{-1}}{nh_2^2})^{1/2}\}\bigg].
\end{eqnarray*}
Consequently, by letting $h_l\to 0$ for $l=1$ and $l=2$, we have
\begin{eqnarray*}
	\sum_{i^*\in\mS_{11}} a_{ii^*} X_{i^*1}&=&I(D_i=\mD_2)\sum_{i^*\in\mS_{11}} a_{ii^*} X_{i^*1}\\
	&=&I(D_i=\mD_2)|\mS_{11}|\bigg[P(D=\mathcal{D}_1|Y_i)E(X_{1}|Y_i)f(Y_i)\\&&+P(D=\mathcal{D}_2|Y_i,X_{i2})E(X_{1}|Y_i,X_{i2})f(Y_i,X_{i2})\\&&+P(D=\mathcal{D}_3|Y_i,X_{i2})E(X_{1}|Y_i,X_{i2})f(Y_i,X_{i2})\bigg]+O_p(h_1^3)+O_p(h_2^3).
\end{eqnarray*}

We subsequently consider the second component $\sum_{i^*\in\mS_{11}} a_{ii^*}$, which is
\begin{eqnarray*}
	\sum_{i^*\in\mS_{11}} a_{ii^*}&=&I(D_i=\mD_2)\sum_{i^*\in\mS_{11}} a_{ii^*} \\
	&=&I(D_i=\mD_2)\bigg[\sum_{i^*\in\mS_{11}}I(D_{i^*}=\mathcal{D}_1)K_{h_1}(Y_i-Y_{i^*})\\
&+&\sum_{i^*\in\mS_{11}}I(D_{i^*}=\mathcal{D}_2)K_{h_1}(Y_i-Y_{i^*})K_{h_2}(X_{i2}-X_{i^*2})\\
	&+&\sum_{i^*\in\mS_{11}}I(D_{i^*}=\mathcal{D}_3)K_{h_1}(Y_i-Y_{i^*})K_{h_2}(X_{i2}-X_{i^*2})\bigg].
\end{eqnarray*}
After algebraic simplification, we obtain that
\begin{eqnarray*}
	&&\sum_{i^*\in\mS_{11}}I(D_{i^*}=\mathcal{D}_1)K_{h_1}(Y_i-Y_{i^*})\\&=&|\mS_{11}|\bigg[P(D=\mathcal{D}_1|Y_i)f(Y_i)+h^3_1\mu_3\frac{\partial [P(D=\mathcal{D}_1|Y_i)f(Y_i)] }{\partial Y_i}+O_p\{h_1^4+(\frac{\log h_1^{-1}}{nh_1^2})^{1/2}\}\bigg], \\\\
	&&\sum_{i^*\in\mS_{11}}I(D_{i^*}=\mathcal{D}_2)K_{h_1}(Y_i-Y_{i^*})K_{h_2}(X_{i2}-X_{i^*2})\\
	&=&|\mS_{11}|\bigg[P(D=\mathcal{D}_2|Y_i,X_{i2})f(Y_i,X_{i2})
	+h^3_1\mu_3\frac{\partial^3 [P(D=\mathcal{D}_2|Y_i,X_{i2})f(Y_i,X_{i2})] }{\partial Y_i^3}\\
	&+&h^3_2\mu_3\frac{\partial^3 [P(D=\mathcal{D}_2|Y_i,X_{i2})f(Y_i,X_{i2})] }{\partial X_{i2}^3}+O_p\{h_1^4+(\frac{\log h_1^{-1}}{nh_1^2})^{1/2}\}+O_p\{h_2^4+(\frac{\log h_2^{-1}}{nh_2^2})^{1/2}\}\bigg],
\end{eqnarray*}
and
\begin{eqnarray*}
	&&	\sum_{i^*\in\mS_{11}}I(D_{i^*}=\mathcal{D}_3)K_{h_1}(Y_i-Y_{i^*})K_{h_2}(X_{i2}-X_{i^*2})\\
	&=&|\mS_{11}|\bigg[P(D=\mathcal{D}_3|Y_i,X_{i2})f(Y_i,X_{i2})+h^3_1\mu_3\frac{\partial^3 [P(D=\mathcal{D}_3|Y_i,X_{i2})f(Y_i,X_{i2})] }{\partial Y_i^3}\\
	&+&h^3_2\mu_3\frac{\partial^3 [P(D=\mathcal{D}_3|Y_i,X_{i2})f(Y_i,X_{i2})] }{\partial X_{i2}^3}+O_p\{h_1^4+(\frac{\log h_1^{-1}}{nh_1^2})^{1/2}\}+O_p\{h_2^4+(\frac{\log h_2^{-1}}{nh_2^2})^{1/2}\}\bigg].
\end{eqnarray*}
The above results imply that
\begin{eqnarray*}
	\sum_{i^*\in\mS_{11}} a_{ii^*}&=&I(D_i=\mD_2)\sum_{i^*\in\mS_{11}} a_{ii^*} \\
	&=&I(D_i=\mD_2)|\mS_{11}|\bigg[P(D=\mathcal{D}_1|Y_i)f(Y_i)+P(D=\mathcal{D}_2|Y_i,X_{i2})f(Y_i,X_{i2})\\&&+P(D=\mathcal{D}_3|Y_i,X_{i2})f(Y_i,X_{i2})+O_p(h_1^3)+O_p(h_2^3)\bigg].
\end{eqnarray*}

By $nh_l^6\to 0$ for $l=1$ and $2$, in conjunction with the results obtained from the two components,  we then have
\begin{eqnarray*}
	J_{1i}&=&\frac{\sum_{i^*\in\mS_{11}} a_{ii^*} X_{i^*1}}{\sum_{i^*\in\mS_{11}} a_{ii^*}}\{1+o_p(1)\}-\frac{\sum_{i^*\in\mS_{11}} a_{ii^*}X_{i1}}{\sum_{i^*\in\mS_{11}} a_{ii^*}}\\
	&=&\frac{P(D=\mathcal{D}_1|Y_i)f(Y_i)[E(X_1|Y_i)\{1+o_p(1)\}-X_{i1}]}{P(D=\mathcal{D}_1|Y_i)f(Y_i)+P(D=\mathcal{D}_2|Y_i,X_{i2})f(Y_i,X_{i2})+P(D=\mathcal{D}_3|Y_i,X_{i2})f(Y_i,X_{i2})}\\
	&&+\frac{P(D=\mathcal{D}_2|Y_i,X_{i2})f(Y_i,X_{i2})[E(X_1|Y_i,X_{i2})\{1+o_p(1)\}-X_{i1}]}{P(D=\mathcal{D}_1|Y_i)f(Y_i)+P(D=\mathcal{D}_2|Y_i,X_{i2})f(Y_i,X_{i2})+P(D=\mathcal{D}_3|Y_i,X_{i2})f(Y_i,X_{i2})}\\
	&&+\frac{P(D=\mathcal{D}_3|Y_i,X_{i2})f(Y_i,X_{i2})[E(X_1|Y_i,X_{i2})\{1+o_p(1)\}-X_{i1}]}{P(D=\mathcal{D}_1|Y_i)f(Y_i)+P(D=\mathcal{D}_2|Y_i,X_{i2})f(Y_i,X_{i2})+P(D=\mathcal{D}_3|Y_i,X_{i2})f(Y_i,X_{i2})}\\
	&&+O_p(h_1^3)+O_p(h_2^3)\\
	&=:&J_{1i1}+J_{1i2}+J_{1i3}+o_p(n^{-1/2}).
\end{eqnarray*}

We next consider $J_{2i}$ in (A.1) for $i\in \mS_{0j}$.
By Conditions (C1) and (C4), we have that $\|C_{\mS_{01}}\|<\infty$.
This, together with the result of $\|\mathcal{W}\|_{\max}=O_p((nh^{d_0})^{-1})$, leads to
 $\max_{i\in \mS_{01}}|J_{2i}|=O_p((nh^{d_0})^{-1})$. Accordingly,
from the result of $J_{1i}$ given above and then employing Condition (C3),  we further have
\begin{eqnarray*}
	&&\bigg|\frac{1}{|\mS_{01}|}\sum_{i\in \mS_{01}}(\hat{X}_{i1}-X_{i1})\bigg|\\
	&=&\bigg|\frac{1}{|\mS_{01}|}\sum_{i\in \mS_{01}}J_{1i}+o_p(n^{-1/2})\bigg|\\
	&=&\bigg|\frac{1}{|\mS_{01}|}\sum_{i\in \mS_{01}}(J_{1i1}+J_{1i2}+J_{1i3})+o_p(|\mS_{01}|^{-1/2})\bigg|\\
	&\leq& \bigg|\frac{1}{|\mS_{01}|}\sum_{i\in \mS_{01}}J_{1i1}+o_p(|\mS_{01}|^{-1/2})\bigg|+
	\bigg|\frac{1}{|\mS_{01}|}\sum_{i\in \mS_{01}}J_{1i2}+o_p(|\mS_{01}|^{-1/2})\bigg|\\
&&+\bigg|\frac{1}{|\mS_{01}|}\sum_{i\in \mS_{01}}J_{1i3}+o_p(|\mS_{01}|^{-1/2})\bigg|=:J_3+J_4+J_5.
\end{eqnarray*}
By Conditions (C1) and (C6), it can be shown that $E(J_{1i1})^2<\infty$,  $E(J_{1i2})^2<\infty$ and $E(J_{1i3})^2<\infty$ as well as that  there exist positive constants $c_k$, $k=1,\cdots,6$, such that
$c_1<P(D=\mathcal{D}_1|Y_i)f(Y_i)<c_2$, $c_3<P(D=\mathcal{D}_2|Y_i,X_{i2})f(Y_i,X_{i2})<c_4$ and $c_5<P(D=\mathcal{D}_3|Y_i,X_{i2})$
$f(Y_i,X_{i2})<c_6$. For any $i\in\mS_{01}$,
define \[a_i=\frac{P(D=\mathcal{D}_1|Y_i)f(Y_i)}{{P(D=\mathcal{D}_1|Y_i)f(Y_i)+P(D=\mathcal{D}_2|Y_i,X_{i2})f(Y_i,X_{i2})+P(D=\mathcal{D}_3|Y_i,X_{i2})f(Y_i,X_{i2})}},\]
\[b_i=\frac{P(D=\mathcal{D}_2|Y_i, X_{i2})f(Y_i)}{{P(D=\mathcal{D}_1|Y_i)f(Y_i)+P(D=\mathcal{D}_2|Y_i,X_{i2})f(Y_i,X_{i2})+P(D=\mathcal{D}_3|Y_i,X_{i2})f(Y_i,X_{i2})}},\]
and
\[c_i=\frac{P(D=\mathcal{D}_3|Y_i, X_{i2})f(Y_i)}{{P(D=\mathcal{D}_1|Y_i)f(Y_i)+P(D=\mathcal{D}_2|Y_i,X_{i2})f(Y_i,X_{i2})+P(D=\mathcal{D}_3|Y_i,X_{i2})f(Y_i,X_{i2})}}.\]
Then, $a_i$, $b_i$ and $c_i$ are all bounded.
Accordingly,
we can show that there exist positive constants $c_7,c_8$ and $c_9$, such that
\begin{eqnarray*}
	&&J_3=\bigg|\frac{1}{|\mS_{01}|}\sum_{i\in \mS_{01}}J_{1i1}+o(|\mS_{01}|^{-1/2})\bigg|\leq c_7 \bigg|\frac{1}{|\mS_{01}|}\sum_{i\in \mS_{01}}a_i\{E(X_{1}|Y_i)-X_{i1}\}\bigg|=:J_6,\\
	&&J_4=\bigg|\frac{1}{|\mS_{01}|}\sum_{i\in \mS_{01}}J_{1i2}+o(|\mS_{01}|^{-1/2})\bigg|\leq c_8 \bigg|\frac{1}{|\mS_{01}|}\sum_{i\in \mS_{01}}b_i\{E(X_{1}|Y_i,X_{i2})-X_{i1}\}\bigg|=:J_7, \mbox{~and~}\\
	&&J_5=\bigg|\frac{1}{|\mS_{01}|}\sum_{i\in \mS_{01}}J_{1i3}+o(|\mS_{01}|^{-1/2})\bigg|\leq c_9 \bigg|\frac{1}{|\mS_{01}|}\sum_{i\in \mS_{01}}c_i\{E(X_{1}|Y_i,X_{i2})-X_{i1}\}\bigg|=:J_8.
\end{eqnarray*}
Using the fact that $a_i\{E(X_{1}|Y_i)-X_{i1}\}$, $b_i\{E(X_{1}|Y_i,X_{i2})-X_{i1}\}$ and $c_i\{E(X_{1}|Y_i,X_{i2})-X_{i1}\}$ are independent sequences
for $i\in \mS_{01}$  and applying
Bernstein inequalities and Condition (C1), we have
\begin{eqnarray*}
	&&P\bigg(\bigg|\frac{1}{|\mS_{01}|}\sum_{i\in \mS_{01}}(\hat{X}_{i1}-X_{i1})\bigg|>\frac{\delta}{n}\bigg)\\
	&\leq&P\bigg(J_3>\frac{\delta}{3n}\bigg)+P\bigg(J_4>\frac{\delta}{3n}\bigg)+P\bigg(J_5>\frac{\delta}{3n}\bigg)\\
	&\leq&P\bigg(J_6>\frac{\delta}{3n}\bigg)+P\bigg(J_7>\frac{\delta}{3n}\bigg)+P\bigg(J_8>\frac{\delta}{3n}\bigg)\\
	&\leq& 2\exp\bigg(-\frac{1}{2}\frac{\delta^2}{c_{10}n+c_{11}\delta}\bigg)+2\exp\bigg(-\frac{1}{2}\frac{\delta^2}{c_{12}n+c_{13}\delta}\bigg)+2\exp\bigg(-\frac{1}{2}\frac{\delta^2}{c_{14}n+c_{15}\delta}\bigg)\\
	&\leq& 6\exp\bigg(-\frac{1}{2}\frac{\delta^2}{C_1n+C_2\delta}\bigg),
\end{eqnarray*}
where $C_1=\max\{c_{10},c_{12},c_{14}\}$, $C_2=\max\{c_{11},c_{13},c_{15}\}$, and
$c_k$ for $k=10,\cdots,15$ are positive constants, which completes the entire proof.

\scsubsection{Appendix C: Additional Simulation Studies}

In this section, we consider six settings for additional simulation studies to examine the robustness of SSI against the non-normality of
covariates, the selection of tuning parameters and different missing mechanisms.
The six settings are:
(I) the covariate $X_i$s being generated from a multivariate exponential distribution;
(II) the selection
of the tuning parameter $\tau$;
(III) an AR(1)  covariance matrix setting with $|\rho|<1$;
(IV) the ``non-ignorable missing data'' setting modified from Kim and Yu (2011);
(V) the ``non-ignorable missing data'' setting suggested by an anonymous referee; and
(VI) the ``missing data with no pattern'' setting.
The detailed  settings and their corresponding simulation  results are given below.

\noindent\textbf{Setting I: Non-normal Covariates.}
Let $Z_i\in\mR^{p-2}$ with $p=10$  be
iid generated from a standard exponential distribution for $i=1,\cdots,n$.
Consider
the continuous covariates $X_{ib}=\Sigma_X^{1/2} Z_i$,
where  $\Sigma_X=(\sigma_{x,j_1j_2})$, and
$\sigma_{x,j_1j_2}=1$ for $j_1=j_2$, and $\sigma_{x,j_1j_2}=\rho<1$ otherwise.
The rest of the setting is the same as in Section 3.1 of the manuscript. The simulation results  in
Tables S1--S2 show similar patterns to those in Tables 3--4 of the paper under normal covariates. Hence our method performs well under a non-normal distribution.

\noindent\textbf{Setting II: Tuning Parameter Selection.}
To select the tuning parameters, we set the range of $\tau$ from 0 to 2 in Section 3.1 of the paper. To assess the sensitivity of the tuning parameter to the selected range, we
consider two additional ranges: 0 to 5 and 0 to 10.
The rest of the setting is the same as in Section 3.1 of the manuscript.
Tables S3--S4 show similar findings to those in Tables 3--4 of the manuscript
as $\tau$ ranges from 0 to 2. Since the range 0 to 10 yields similar results, we do not present them here.
In sum, our method is not sensitive to the selection range of $\tau$.

\noindent\textbf{Setting III:  Covariance Matrix with AR(1) Setting.}
Motivated by an anonymous referee's comments, we generate the $X_i$s by adopting the AR(1) covariance matrix setting from Pourahmadi (2013, p.54). The structure of AR(1) matrix is as follows:
$\Sigma_X=(\sigma_{x,j_1j_2})$ and
$\sigma_{x,j_1j_2}=\rho^{|j_1-j_2|}$, where $|\rho|<1$.
This setting assures $\Sigma_X$ positive definite even when $\rho$
is negative with $-1<\rho<0$.
We consider $\rho=$ -0.25, -0.5 and -0.75 in this study, and the rest of the simulation setting is the same as in Section 3.1 of the paper.

The simulation results are summarized in
Table S5, and they exhibit a similar pattern to those in Table 4 of the paper.
Hence, our method still performs satisfactorily when $\rho$ is negative in this setting.

\noindent\textbf{Setting IV:  Non-ignorable Missing Data I.}  Under MNAR, we consider
non-ignorable missing data, as motivated by Kim and Yu (2011). We name it MNAR2 hereafter.
Specifically, for each of $i=1,\cdots, n$
and  $j=1,\cdots, p-1$,
the $D_{ij}$  is independently generated from a binomial distribution with probability
$\{1+\mbox{exp}(-\epsilon_i^2-\alpha_3)\}^{-1}$, where $\epsilon_i$ is quadratic and $\alpha_3$ is selected to control the
missing rate at 50\%.
The rest of the setting is the same as in Section 3.1 of the paper.
Tables S6--S7 exhibit similar patterns to those in Tables 3--4 of the paper.
Hence, our method still performs well under this ``non-ignorable missing data'' setting.

\noindent\textbf{Setting V:  Non-ignorable Missing Data II.} Under MNAR, we consider another
non-ignorable missing data suggested by an anonymous referee. We name it MNAR3 hereafter.
Specifically, for each of $i=1,\cdots, n$
and  $j=1,\cdots, p$,
the $D_{ij}$  is independently generated from a binomial distribution with probability
$\big\{1+\exp(-\alpha_4-0.5 X_{ij}-0.5 X_{ij}^2-0.5 X_{ij}X_{i,j-1})\big\}^{-1}$, where $X_{i0}=0$ and $\alpha_4$ is selected to control the
missing rate at 50\%. Tables S8--S9 exhibit similar patterns to those in Tables 3--4 of the paper.
Accordingly, our method  performs well under this  ``non-ignorable missing data'' setting.

\noindent\textbf{Setting VI: Missing Data With No Pattern.}
We consider missing data that are generated without specifying any missing patterns.
To this end, we conduct the following three scenarios.

(a)  Under MCAR, for $i=1,\cdots, n$
and  $j=1,\cdots, p-1$, the $D_{ij}$ are independently generated from  binomial distributions that do not relate to any
covariates and errors. In addition, the missing proportion in covariates is approximately 50\%.

(b) Under MAR,  for  $i=1,\cdots, n$
and  $j=1,\cdots, p-1$, the
$D_{ij}$ are independently generated from a binomial distribution with probability
 $\{1+\mbox{exp}(-0.5X_{ip}-\alpha_1)\}^{-1}$, which is related to the observed
covariate $X_{ip}$. Furthermore, $\alpha_1$ is chosen to control the missing rate
at approximately 50\%.

(c) Under MNAR, for $i=1,\cdots, n$
and  $j=1,\cdots, p-1$,
the $D_{ij}$  are independently generated from a binomial distribution with probability
$\{1+\mbox{exp}(-\epsilon_i-\alpha_2)\}^{-1}$, where $\alpha_2$ is selected to control the
missing rate at  approximately 50\%.

The rest of the setting for each of the above three scenarios is the same as  in Section 3.1 of the paper.
Tables S10--S11 indicate that the simulation results are similar to those in Tables 3--4 of the paper. As a result, SSI performs well for no missing patterns.

\scsubsection{Appendix D: Sequentially Semi-Supervised Imputation (SSSI)}

In this section, we present simulation results for SSSI, which are discussed in the concluding remark sof the manuscript.
The simulation setting is the same as in Section 3.1 of the paper. For comparing with
SSI$_1$ and SSI$_2$, we denote their corresponding SSSI by SSSI$_1$ and SSSI$_2$.
The simulation results are summarized in Tables S12--S13 with $m=10$.
We find that SSSI$_1$ and SSSI$_2$ are very slightly better than SSI$_1$ and SSI$_2$, respectively.
However, SSSI is a sequential algorithm and obtaining its theoretical properties is a challenging task.  Hence we consider it for future research in our concluding remarks.

\scsection{Additional Reference}

\begin{description}
\newcommand{\enquote}[1]{``#1''}
\expandafter\ifx\csname natexlab\endcsname\relax\def\natexlab#1{#1}\fi

\bibitem[{Pourahmadi(2013)}]{Pourahmadi:2013}
Pourahmadi, M. (2013).
 \textit{High-Dimensional Covariance Estimation,}   New York: John Wiley.

 \end{description}

\begin{landscape}
\begin{table}[htbp!]
\begin{center}
\caption
{The simulation results of IA, EA and PA under the
three coefficients of determination ($R^2$=0.3, 0.6 and 0.9) and three missing mechanisms (MCAR, MAR, and MNAR),
and they are obtained by averaging over three correlations ($\rho=0.25$, 0.5 and 0.75) and three sample sizes ($n=500$, 1,000 and 2,000) for Setting I.
The value in parentheses is the averaged standard error.}
\vspace{0.28 cm}
\begin{tabular}{c|c|ccc|ccc|ccc}
\hline
 &  &  \multicolumn{3}{|c}{MCAR} & \multicolumn{3}{|c}{MAR} & \multicolumn{3}{|c}{MNAR} \\
$R^2$ & Methods & IA & EA  &  PA & IA & EA  &  PA & IA & EA  &  PA \\
\hline
 0.3     & KNN & 1.99(0.842) & 0.77(0.061) & 1.02(0.097) & 2.03(0.904) & 0.85(0.066) & 0.96(0.099) & 2.04(0.911) & 0.85(0.069) & 0.98(0.124) \\
         & RF & 1.91(0.761) & 0.70(0.042) & 0.87(0.081) & 1.95(0.753) & 0.71(0.047) & 0.89(0.091) & 1.94(0.747) & 0.70(0.044) & 0.89(0.086) \\
         & MI & 2.62(0.833) & 0.76(0.028) & 0.91(0.087) & 2.59(0.839) & 0.76(0.029) & 0.93(0.085) & 2.61(0.849) & 0.76(0.027) & 0.92(0.088) \\
         & RI & 2.02(0.688) & 0.70(0.205) & 1.16(0.544) & 2.05(0.709) & 2.31(2.846) & 3.26(2.423) & 2.24(0.721) & 1.23(1.010) & 2.43(2.144) \\
         & MLE & - & 0.74(1.219) & - & - & 0.65(0.085) & - & - & 0.73(0.581) & - \\
         & SSI$_1$ & 1.78(0.782) & 0.55(0.054) & 0.90(0.088) & 1.79(0.776) & 0.56(0.062) & 0.92(0.093) & 1.79(0.781) & 0.56(0.064) & 0.91(0.092) \\
         & SSI$_2$ & 1.78(0.787) & 0.62(0.051) & 0.85(0.083) & 1.79(0.781) & 0.63(0.055) & 0.86(0.083) & 1.79(0.787) & 0.63(0.054) & 0.86(0.084) \\
         \hline
 0.6     & KNN & 1.99(0.822) & 0.72(0.059) & 0.80(0.091) & 2.02(0.824) & 0.74(0.064) & 0.81(0.102) & 2.02(0.827) & 0.79(0.071) & 0.81(0.114) \\
         & RF & 1.90(0.744) & 0.62(0.048) & 0.73(0.087) & 1.95(0.752) & 0.63(0.052) & 0.75(0.092) & 1.95(0.753) & 0.63(0.052) & 0.75(0.091) \\
         & MI & 2.62(0.861) & 0.70(0.027) & 0.81(0.093) & 2.61(0.832) & 0.70(0.026) & 0.82(0.093) & 2.60(0.851) & 0.70(0.026) & 0.81(0.094) \\
         & RI & 2.01(0.677) & 0.60(0.157) & 1.02(0.558) & 2.05(0.706) & 0.96(0.723) & 1.92(1.812) & 2.39(0.71) & 1.39(1.493) & 2.57(2.198) \\
         & MLE & - & 0.59(1.472) & - & - & 0.56(0.127) & - & - & 0.57(0.260) & - \\
         & SSI$_1$ & 1.79(0.771) & 0.50(0.045) & 0.69(0.074) & 1.79(0.777) & 0.50(0.044) & 0.70(0.078) & 1.79(0.773) & 0.50(0.047) & 0.70(0.075) \\
         & SSI$_2$ & 1.77(0.773) & 0.53(0.056) & 0.67(0.074) & 1.79(0.785) & 0.53(0.054) & 0.68(0.078) & 1.78(0.774) & 0.54(0.056) & 0.68(0.075) \\
         \hline
 0.9     & KNN & 1.98(0.821) & 0.64(0.058) & 0.67(0.084) & 2.02(0.812) & 0.62(0.061) & 0.68(0.134) & 2.01(0.810) & 0.60(0.064) & 0.70(0.112) \\
         & RF & 1.92(0.742) & 0.57(0.051) & 0.58(0.092) & 1.94(0.755) & 0.58(0.053) & 0.61(0.12) & 1.96(0.759) & 0.58(0.056) & 0.62(0.102) \\
         & MI & 2.61(0.855) & 0.65(0.026) & 0.71(0.098) & 2.61(0.837) & 0.66(0.028) & 0.71(0.11) & 2.60(0.841) & 0.66(0.025) & 0.71(0.098) \\
         & RI & 2.02(0.676) & 0.51(0.103) & 0.87(0.314) & 2.04(0.706) & 0.61(0.309) & 1.16(0.946) & 2.18(0.712) & 1.61(1.718) & 2.54(2.083) \\
         & MLE & - & 0.44(0.043) & - & - & 0.47(0.117) & - & - & 0.49(0.088) & - \\
         & SSI$_1$ & 1.79(0.770) & 0.49(0.041) & 0.50(0.073) & 1.79(0.778) & 0.49(0.039) & 0.52(0.077) & 1.80(0.777) & 0.49(0.04) & 0.52(0.077) \\
         & SSI$_2$ & 1.79(0.774) & 0.48(0.054) & 0.50(0.073) & 1.79(0.785) & 0.47(0.054) & 0.51(0.078) & 1.80(0.789) & 0.48(0.056) & 0.51(0.077) \\
         \hline
\end{tabular}
\end{center}
\end{table}
\end{landscape}

\begin{landscape}
\begin{table}[htbp!]
\begin{center}
\caption
{The simulation results of IA, EA and PA under the three correlations ($\rho=0.25$, 0.5 and 0.75) and three missing mechanisms (MCAR, MAR, and MNAR), and they are obtained by averaging over the
three coefficients of determination ($R^2$=0.3, 0.6 and 0.9) and three sample sizes ($n=500$, 1,000 and 2,000) for Setting I. The value in parentheses is the averaged standard error.
Note that FIME is only applicable for parameter estimation; hence, we display its IA and PA as ``-".}
\vspace{0.28 cm}
\begin{tabular}{c|c|ccc|ccc|ccc}
\hline
 &  &  \multicolumn{3}{|c}{MCAR} & \multicolumn{3}{|c}{MAR} & \multicolumn{3}{|c}{MNAR} \\
$\rho$ & Methods & IA & EA  &  PA & IA & EA  &  PA & IA & EA  &  PA \\
\hline
 0.25    & KNN & 1.32(0.115) & 0.70(0.071) & 0.82(0.151) & 1.46(0.115) & 0.71(0.080) & 0.86(0.162) & 1.44(0.112) & 0.79(0.082) & 0.91(0.144) \\
         & RF & 1.22(0.103) & 0.62(0.067) & 0.75(0.138) & 1.25(0.104) & 0.63(0.072) & 0.78(0.136) & 1.26(0.104) & 0.66(0.071) & 0.79(0.132) \\
         & RI & 1.41(0.148) & 0.59(0.126) & 1.04(0.454) & 1.40(0.133) & 0.82(0.639) & 1.75(1.624) & 1.37(0.125) & 1.71(1.487) & 3.79(2.700) \\
         & FIME & - & 0.53(0.119) & - & - & 0.53(0.085) & - & - & 0.53(0.083) & - \\
         & SSI$_1$ & 1.05(0.088) & 0.49(0.039) & 0.72(0.170) & 1.06(0.083) & 0.49(0.042) & 0.73(0.170) & 1.06(0.081) & 0.49(0.041) & 0.74(0.168) \\
         & SSI$_2$ & 1.05(0.088) & 0.52(0.067) & 0.70(0.156) & 1.05(0.083) & 0.52(0.075) & 0.71(0.151) & 1.05(0.081) & 0.52(0.071) & 0.71(0.151) \\
         \hline
 0.5     & KNN & 1.77(0.142) & 0.65(0.074) & 0.80 (0.152) & 1.83(0.154) & 0.68(0.071) & 0.82(0.153) & 1.80(0.151) & 0.77(0.091) & 0.82(0.167) \\
         & RF & 1.60(0.128) & 0.61(0.070) & 0.75(0.138) & 1.62(0.135) & 0.62(0.070) & 0.76(0.147) & 1.62(0.139) & 0.62(0.071) & 0.77(0.136) \\
         & MI & 2.22(0.193) & 0.69(0.051) & 0.77(0.130) & 2.22(0.200) & 0.70(0.050) & 0.79(0.131) & 2.20(0.197) & 0.70(0.050) & 0.78(0.127) \\
         & RI & 1.72(0.171) & 0.61(0.143) & 1.02(0.486) & 1.72(0.167) & 0.95(1.109) & 1.70(1.676) & 1.69(0.137) & 1.52(1.337) & 2.88(2.203) \\
         & FIME & - & 0.54(0.097) & - & - & 0.55(0.102) & - & - & 0.57(0.185) & - \\
         & SSI$_1$ & 1.47(0.122) & 0.49(0.040) & 0.71(0.172) & 1.48(0.125) & 0.49(0.041) & 0.72(0.179) & 1.48(0.128) & 0.49(0.041) & 0.72(0.170) \\
         & SSI$_2$ & 1.47(0.122) & 0.51(0.070) & 0.69(0.157) & 1.48(0.124) & 0.51(0.074) & 0.70(0.161) & 1.48(0.127) & 0.51(0.075) & 0.70(0.157) \\
         \hline
 0.75    & KNN & 3.23(0.301) & 0.72(0.071) & 0.77(0.176) & 3.32(0.277) & 0.74(0.076) & 0.81(0.179) & 3.10(0.269) & 0.073(0.080) & 0.80(0.162) \\
         & RF & 2.92(0.260) & 0.67(0.065) & 0.69(0.155) & 2.96(0.256) & 0.67(0.067) & 0.71(0.158) & 2.96(0.259) & 0.67(0.063) & 0.71(0.149) \\
         & MI & 3.75(0.319) & 0.71(0.048) & 0.79(0.124) & 3.72(0.302) & 0.72(0.046) & 0.80(0.127) & 3.73(0.319) & 0.72(0.044) & 0.80(0.123) \\
         & RI & 2.91(0.276) & 0.61(0.243) & 0.99(0.550) & 2.96(0.285) & 1.02(1.526) & 1.63(1.714) & 2.94(0.269) & 1.11(1.185) & 1.85(1.625) \\
         & FIME & - & 0.71(1.910) & - & - & 0.60(0.185) & - & - & 0.69(0.621) & - \\
         & SSI$_1$ & 2.82(0.249) & 0.56(0.045) & 0.67(0.192) & 2.83(0.249) & 0.57(0.050) & 0.68(0.194) & 2.83(0.248) & 0.57(0.053) & 0.67(0.184) \\
         & SSI$_2$ & 2.83(0.250) & 0.61(0.060) & 0.64(0.174) & 2.85(0.249) & 0.61(0.066) & 0.65(0.174) & 2.85(0.250) & 0.61(0.066) & 0.65(0.168) \\
         \hline
\end{tabular}
\end{center}
\end{table}
\end{landscape}

\begin{landscape}
\begin{table}[htbp!]
\begin{center}
\caption
{The simulation results of IA, EA and PA under the
three coefficients of determination ($R^2$=0.3, 0.6 and 0.9) and three missing mechanisms (MCAR, MAR, and MNAR),
and they are obtained by averaging over three correlations ($\rho=0.25$, 0.5 and 0.75) and three sample sizes ($n=500$, 1,000 and 2,000) for $\tau$ ranging from
0 to 5. The value in parentheses is the averaged standard error.}
\vspace{0.28 cm}
\begin{tabular}{c|c|ccc|ccc|ccc}
\hline
 &  &  \multicolumn{3}{|c}{MCAR} & \multicolumn{3}{|c}{MAR} & \multicolumn{3}{|c}{MNAR} \\
$R^2$ & Methods & IA & EA  &  PA & IA & EA  &  PA & IA & EA  &  PA \\
\hline
0.3      & KNN  & 0.79(0.197) & 0.76(0.061) & 0.83(0.098) & 1.02(0.216) & 0.74(0.062) & 0.97(0.124) & 0.79(0.219) & 0.76(0.064) & 0.97(0.094) \\
         & RF   & 0.74(0.228) & 0.78(0.027) & 0.78(0.081) & 0.75(0.234) & 0.78(0.027) & 0.79(0.085) & 0.74(0.233) & 0.78(0.027) & 0.78(0.083) \\
         & MI   & 1.15(0.347) & 0.79(0.024) & 0.80(0.085) & 1.15(0.347) & 0.79(0.025)  & 0.80(0.088) & 1.15(0.351) & 0.79(0.023) & 0.79(0.088) \\
         & RI   & 0.86(0.262) & 0.81(0.141) & 0.97(0.399) & 0.86(0.237) & 1.28(0.959) & 2.12(1.962) & 0.82(0.266) & 0.94(0.498) & 1.39(1.347) \\
         & FIME  & -           & 0.76(0.088) & -            & -            & 0.76(0.048) & -            & -            & 0.76(0.042)  & -       \\
         & SSI$_1$ & 0.68(0.164) & 0.71(0.068) & 0.82(0.084) & 0.68(0.163) & 0.71(0.074) & 0.83(0.083) & 0.68(0.165) & 0.71(0.063) & 0.80(0.086) \\
         & SSI$_2$ & 0.68(0.163) & 0.73(0.049) & 0.77(0.078) & 0.68(0.165) & 0.73(0.051) & 0.78(0.079) & 0.74(0.135) & 0.73(0.052) & 0.77(0.080) \\
         \hline
0.6      & KNN  & 0.79(0.192) & 0.70(0.077) & 0.73(0.096) & 0.86(0.197) & 0.70(0.069) & 0.73(0.094) & 0.79(0.213) & 0.71(0.068) & 0.80(0.091) \\
         & RF   & 0.74(0.229) & 0.73(0.027) & 0.54(0.088) & 0.75(0.234) & 0.73(0.028) & 0.54(0.090) & 0.74(0.235) & 0.73(0.028) & 0.54(0.093) \\
         & MI   & 1.15(0.351) & 0.74(0.022) & 0.57(0.094) & 1.15(0.348) & 0.74(0.022)  & 0.57(0.092) & 1.15(0.349) & 0.74(0.021) & 0.57(0.096) \\
         & RI   & 0.86(0.260)  & 0.73(0.093) & 0.66(0.252)  & 0.88(0.255) & 0.81(0.341) & 0.89(0.894) & 0.80(0.234) & 1.13(0.723) & 1.79(2.039) \\
         & FIME  & -           & 0.69(0.045) & -            & -            & 0.69(0.048)  & -            & -            & 0.69(0.047) & -            \\
         & SSI$_1$ & 0.68(0.164) & 0.66(0.053) & 0.54(0.073) & 0.68(0.166) & 0.66(0.056) & 0.55(0.075) & 0.68(0.165) & 0.66(0.055) & 0.54(0.076) \\
         & SSI$_2$ & 0.68(0.163) & 0.66(0.055) & 0.52(0.074) & 0.68(0.167) & 0.66(0.057) & 0.53(0.078) & 0.68(0.160) & 0.66(0.052) & 0.52(0.075) \\
         \hline
0.9      & KNN  & 0.80(0.192) & 0.66(0.062) & 0.48(0.125) & 0.81(0.197) & 0.76(0.072) & 0.48(0.119) & -            & -            & -            \\
         & RF   & 0.74(0.229) & 0.69(0.028) & 0.29(0.107) & 0.75(0.234) & 0.69(0.027) & 0.31(0.111) & 0.75(0.234) & 0.69(0.027) & 0.31(0.121) \\
         & MI   & 1.15(0.349) & 0.70(0.021) & 0.34(0.121)  & 1.15(0.352)  & 0.70(0.021) & 0.34(0.119) & 1.15(0.345) & 0.71(0.061) & 0.35(0.122)  \\
         & RI   & 0.86(0.261) & 0.67(0.049) & 0.36(0.168) & 0.85(0.264) & 0.66(0.045) & 0.34(0.142) & 0.75(0.222) & 1.92(1.709) & 2.63(2.486) \\
         & FIME  & -            & 0.64(0.044)  & -            & -            & 0.64(0.043) & -            & -            & 0.64(0.045)  & -            \\
         & SSI$_1$ & 0.68(0.166) & 0.65(0.039) & 0.28(0.087) & 0.68(0.167) & 0.65(0.040) & 0.28(0.090) & 0.68(0.167) & 0.65(0.039) & 0.29(0.091) \\
         & SSI$_2$ & 0.68(0.163) & 0.62(0.059) & 0.26(0.079) & 0.68(0.159) & 0.62(0.053) & 0.27(0.083) & 0.68(0.160) & 0.62(0.057) & 0.27(0.083) \\
         \hline
\end{tabular}
\end{center}
\end{table}
\end{landscape}

\begin{landscape}
\begin{table}[htbp!]
\begin{center}
\caption
{The simulation results of IA, EA and PA under the three correlations ($\rho=0.25$, 0.5 and 0.75)
and three missing mechanisms (MCAR, MAR, and MNAR), and they are obtained by averaging over the
three coefficients of determination ($R^2$=0.3, 0.6 and 0.9) and three sample sizes ($n=500$, 1,000 and 2,000) for $\tau$ ranging from 0 to 5. }
\vspace{0.28 cm}
\begin{tabular}{c|c|ccc|ccc|ccc}
\hline
 &  &  \multicolumn{3}{|c}{MCAR} & \multicolumn{3}{|c}{MAR} & \multicolumn{3}{|c}{MNAR} \\
$\rho$ & Methods & IA & EA  &  PA & IA & EA  &  PA & IA & EA  &  PA \\
\hline
  0.25   & KNN  & 1.00(0.066) & 0.67(0.075) & 0.79(0.223) & 1.03(0.089) & 0.84(0.072) & 0.74(0.239) & 1.04(0.110)   & 0.80(0.074) & 0.85(0.208) \\
         & RF   & 1.01(0.056) & 0.71(0.048) & 0.62(0.178) & 1.02(0.060) & 0.71(0.051) & 0.64(0.178) & 1.02(0.059) & 0.71(0.048) & 0.63(0.175) \\
         & MI   & 1.54(0.095) & 0.73(0.042) & 0.67(0.163) & 1.54(0.094) & 0.73(0.045) & 0.67(0.165) & 1.54(0.094) & 0.73(0.043)  & 0.67(0.161) \\
         & RI   & 1.15(0.101) & 0.70(0.104) & 0.75(0.372) & 1.14(0.103) & 0.77(0.318) & 1.06(1.265) & 1.13(0.098) & 1.28(0.829) & 2.73(2.219) \\
         & FIME  & -           & 0.65(0.078)  & -            & -            & 0.65(0.061) & -            & -            & 0.65(0.067) & -            \\
         & SSI$_1$ & 0.86(0.039) & 0.62(0.034) & 0.61(0.203) & 0.88(0.039) & 0.61(0.032) & 0.62(0.201) & 0.86(0.039) & 0.62(0.035) & 0.61(0.195) \\
         & SSI$_2$ & 0.86(0.040) & 0.61(0.061) & 0.59(0.195) & 0.86(0.040) & 0.61(0.058) & 0.59(0.193) & 0.85(0.038) & 0.61(0.062) & 0.59(0.190) \\
         \hline
  0.5    & KNN  & 0.81(0.072)   & 0.76(0.065) & 0.72(0.411) & 0.88(0.142) & 0.89(0.054)  & 0.75(0.285) & 0.86(0.111)  & 0.73(0.061)  & 0.86(0.298) \\
         & RF   & 0.75(0.046) & 0.73(0.043) & 0.54(0.211) & 0.76(0.046) & 0.74(0.043) & 0.53(0.211) & 0.76(0.044) & 0.74(0.041) & 0.54(0.208) \\
         & MI   & 1.19(0.082) & 0.75(0.037) & 0.56(0.199) & 1.19(0.089) & 0.75(0.039) & 0.57(0.201) & 1.19(0.087)  & 0.75(0.036) & 0.57(0.195) \\
         & RI   & 0.88(0.095) & 0.74(0.099) & 0.65(0.369) & 0.89(0.098) & 0.86(0.486) & 1.02(1.448) & 0.88(0.085) & 1.01(0.587)  & 1.53(1.661) \\
         & FIME  & -            & 0.70(0.053) & -            & -            & 0.70(0.057) & -            & -            & 0.71(0.055) & -            \\
         & SSI$_1$ & 0.71(0.036) & 0.68(0.036) & 0.54(0.230) & 0.71(0.035) & 0.68(0.037) & 0.55(0.232) & 0.71(0.035) & 0.68(0.038) & 0.54(0.220) \\
         & SSI$_2$ & 0.71(0.037) & 0.68(0.052) & 0.51(0.216) & 0.72(0.037) & 0.67(0.054) & 0.52(0.217) & 0.73(0.055) & 0.67(0.054) & 0.52(0.215) \\
         \hline
  0.75   & KNN  & 0.57(0.076) & 0.84(0.059) & 0.66(0.430)  & 0.68(0.217)  & 0.92(0.069) & 0.59(0.301) & 0.58(0.123) & 0.86(0.062) & 0.68(0.285) \\
         & RF   & 0.49(0.028) & 0.75(0.04) & 0.46(0.235) & 0.49(0.029) & 0.75(0.039) & 0.48(0.235) & 0.50(0.027) & 0.75(0.039) & 0.49(0.233) \\
         & MI   & 0.71(0.061) & 0.76(0.036)  & 0.48(0.227) & 0.72(0.063) & 0.76(0.038)  & 0.48(0.226) & 0.72(0.061) & 0.76(0.035)  & 0.48(0.219) \\
         & RI   & 0.55(0.076) & 0.78(0.135) & 0.58(0.387) & 0.57(0.087) & 0.98(0.808) & 0.91(1.27)  & 0.55(0.083) & 1.08(0.948) & 1.16(1.353) \\
         & FIME  & -            & 0.74(0.079) & -            & -            & 0.74(0.051) & -            & -            & 0.74(0.048) & -            \\
         & SSI$_1$ & 0.47(0.024) & 0.73(0.048) & 0.49(0.256) & 0.47(0.023) & 0.74(0.051) & 0.49(0.257) & 0.46(0.025) & 0.73(0.043) & 0.48(0.244) \\
         & SSI$_2$ & 0.47(0.025) & 0.73(0.041) & 0.46(0.237) & 0.47(0.025) & 0.73(0.046) & 0.46(0.236) & 0.52(0.102) & 0.72(0.045) & 0.46(0.235) \\
         \hline
\end{tabular}
\end{center}
\end{table}
\end{landscape}

\begin{landscape}
\begin{table}[htbp!]
\begin{center}
\caption
{The simulation results of IA, EA and PA under the three correlations ($\rho=-0.25$, -0.5 and -0.75) and three missing mechanisms (MCAR, MAR, and MNAR), and they are obtained by averaging over the
three coefficients of determination ($R^2$=0.3, 0.6 and 0.9) and three sample sizes ($n=500$, 1,000 and 2,000) for Setting III. The value in parentheses is the averaged standard error.
Note that FIME is only applicable for parameter estimation; hence, we display its IA and PA as ``-".}
\vspace{0.28 cm}
\begin{tabular}{c|c|ccc|ccc|ccc}
\hline
 &  &  \multicolumn{3}{|c}{MCAR} & \multicolumn{3}{|c}{MAR} & \multicolumn{3}{|c}{MNAR} \\
$\rho$ & Methods & IA & EA  &  PA & IA & EA  &  PA & IA & EA  &  PA \\
\hline
 -0.25    & KNN & 1.25(0.121) & 0.80(0.081) & 0.87(0.201) & 1.64(0.132) & 0.81(0.106) & 0.90(0.177) & 1.60(0.122) & 0.87(0.119) & 1.04(0.176) \\
         & RF & 1.11(0.063)	 & 0.52(0.084) & 0.79(0.124) & 1.13(0.067) & 0.52(0.087) & 0.82(0.127) & 1.14(0.065) & 0.52(0.087) & 0.83(0.123) \\
         & MI & 1.67(0.095)  & 0.66(0.056) & 0.84(0.107) & 1.67(0.099) & 0.67(0.058) & 0.86(0.105) & 1.67(0.103) & 0.67(0.056) & 0.86(0.102) \\
         & RI & 1.24(0.105)	 & 0.48(0.154) & 1.20(0.483) & 1.23(0.096) & 0.68(0.587) & 2.02(1.792) & 1.21(0.099) & 0.99(0.606) & 2.68(2.317) \\
         & FIME & - & 0.47(0.113) & - & - & 0.47(0.095) & - & - & 0.47(0.095) & - \\
         & SSI$_1$ & 0.89(0.040)  & 0.43(0.045) & 0.77(0.155) & 0.89(0.039) & 0.43(0.044) & 0.78(0.155) & 0.89(0.039) & 0.43(0.044) & 0.78(0.152) \\
         & SSI$_2$ & 0.89(0.040)  & 0.46(0.081) & 0.74(0.142) & 0.89(0.039)	& 0.47(0.078) &	0.75(0.137)	& 0.89(0.038) &	0.47(0.081)	& 0.76(0.137) \\
         \hline
 -0.5     & KNN & 1.89(0.165) & 0.79(0.092) & 0.94 (0.167) & 1.90(0.174) & 0.82(0.095) & 0.102(0.177) & 2.01(0.178) & 0.96(0.101) & 0.109(0.182) \\
         & RF & 1.04(0.072)	 & 0.45(0.091) & 0.80(0.126)	& 1.06(0.073) &	0.46(0.094)	& 0.81(0.124) &	1.06(0.074) & 0.46(0.092) &	0.82(0.116) \\
         & MI & 1.52(0.102)	 & 0.62(0.066) & 0.85(0.110)& 1.52(0.097) & 0.63(0.064) & 0.85(0.111) &	1.52(0.101) & 0.64(0.062) &	0.86(0.105) \\
         & RI & 1.09(0.103)	 & 0.41(0.158) & 1.22(0.572) & 1.11(0.094) & 0.65(0.694) &	1.95(1.775) &	1.07(0.085)	& 1.90(2.399)& 2.79(2.246) \\
         & FIME & - & 0.39(0.099) & - & - & 0.39(0.110) & - & - & 0.40(0.109) & - \\
         & SSI$_1$ & 0.86(0.043) & 0.38(0.047) & 0.79(0.158) & 0.86(0.042) & 0.38(0.049) & 0.80(0.152) & 0.86(0.042) & 0.38(0.051) & 0.81(0.148) \\
         & SSI$_2$ & 0.86(0.044) & 0.41(0.083) & 0.77(0.144) & 0.86(0.043) & 0.42(0.087) & 0.77(0.141) & 0.86(0.043) & 0.42(0.082) & 0.77(0.131) \\
         \hline
 -0.75    & KNN & 3.45(0.322) & 0.81(0.078) & 0.89(0.182) & 3.22(0.289) & 0.86(0.096) & 0.97(0.189) & 3.28(0.280) & 0.85(0.087) & 0.96(0.187) \\
         & RF & 0.79(0.073)	& 0.37(0.101) &	0.79(0.121) & 0.80(0.067) &	0.38(0.108) & 0.81(0.121) &	0.81(0.071)	& 0.39(0.103) &	0.81(0.118) \\
         & MI & 1.17(0.096)	& 0.55(0.074) &	0.84(0.108)	& 1.19(0.100) &	0.57(0.081)	& 0.85(0.113) &	1.19(0.103) & 0.57(0.076) &	0.85(0.111) \\
         & RI & 0.87(0.103)	& 0.36(0.186) &	1.15(0.516)	& 0.93(0.097) &	0.87(1.464)	& 1.89(1.733) &	0.93(0.078)	& 1.49(1.513) &	2.93(2.816) \\
         & FIME & - & 0.36(0.149) & - & - & 0.36(0.174) & - & - & 0.36(0.166) & - \\
         & SSI$_1$ & 0.76(0.045) &	0.33(0.056)	& 0.81(0.146) &	0.76(0.044) & 0.33(0.056) &	0.82(0.149)	& 0.76(0.044) &	0.33(0.055)	& 0.82(0.152) \\
         & SSI$_2$ & 0.78(0.065) & 	0.36(0.091)	& 0.78(0.142) &	0.77(0.055)	& 0.36(0.085) &	0.79(0.136)	& 0.78(0.062) &	0.36(0.081)	& 0.78(0.130) \\
         \hline
\end{tabular}
\end{center}
\end{table}
\end{landscape}

\begin{landscape}
\begin{table}[htbp!]
\begin{center}
\caption
{The simulation results of IA, EA and PA under the
three coefficients of determination ($R^2$=0.3, 0.6 and 0.9) and missing mechanism MNAR2,
and they are obtained by averaging over three correlations ($\rho=0.25$, 0.5 and 0.75) and three sample sizes ($n=500$, 1,000 and 2,000) for Setting III.
The value in parentheses is the averaged standard error.
Note that FIME is only applicable for parameter estimation, hence, we display its IA and PA as ``-".}
\vspace{0.28 cm}
\begin{tabular}{c|c|ccc}
\hline
 &  &  \multicolumn{3}{|c}{MNAR2} \\
$R^2$ & Methods & IA & EA  &  PA \\
\hline
0.3      & KNN & 0.81(0.256) & 0.82(0.033) & 0.91(0.102) \\
         & RF & 0.75(0.235) & 0.78(0.028) & 0.79(0.078) \\
         & MI & 1.17(0.350) & 0.79(0.023) & 0.80(0.079) \\
         & RI & 0.87(0.226) & 1.85(3.901) & 2.23(2.235) \\
         & FIME & - & 0.76(0.060) & - \\
         & SSI$_1$ & 0.68(0.156) & 0.69(0.070) & 0.83(0.078) \\
         & SSI$_2$ & 0.69(0.154) & 0.72(0.050) & 0.78(0.073) \\
         \hline
0.6      & KNN & 0.80(0.257) & 0.77(0.031) & 0.61(0.100) \\
         & RF & 0.75(0.236) & 0.73(0.027) & 0.54(0.090) \\
         & MI & 1.15(0.351) & 0.74(0.023) & 0.58(0.099) \\
         & RI & 0.89(0.241) & 1.00(1.184) & 1.21(1.500) \\
         & FIME & - & 0.69(0.047) & - \\
         & SSI$_1$ & 0.68(0.159) & 0.64(0.055) & 0.55(0.074) \\
         & SSI$_2$ & 0.69(0.156) & 0.65(0.052) & 0.52(0.074) \\
         \hline
0.9      & KNN & 0.81(0.255) & 0.73(0.029) & 0.42(0.132) \\
         & RF & 0.75(0.235) & 0.69(0.028) & 0.30(0.113) \\
         & MI & 1.15(0.351) & 0.71(0.020) & 0.35(0.123) \\
         & RI & 0.88(0.247) & 0.70(0.226) & 0.49(0.653) \\
         & FIME & - & 0.64(0.048) & - \\
         & SSI$_1$ & 0.67(0.162) & 0.63(0.042) & 0.28(0.087) \\
         & SSI$_2$ & 0.68(0.155) & 0.61(0.053) & 0.27(0.082) \\
         \hline
\end{tabular}
\end{center}
\end{table}
\end{landscape}

\begin{landscape}
\begin{table}[htbp!]
\begin{center}
\caption
{The simulation results of IA, EA and PA under the three correlations ($\rho=0.25$, 0.5 and 0.75) and missing mechanism MNAR2, and they are obtained by averaging over the
three coefficients of determination ($R^2$=0.3, 0.6 and 0.9) and three sample sizes ($n=500$, 1,000 and 2,000) for Setting III. The value in parentheses is the averaged standard error.
Note that FIME is only applicable for parameter estimation; hence, we display its IA and PA as ``-".}
\vspace{0.28 cm}
\begin{tabular}{c|c|ccc}
\hline
 &  &  \multicolumn{3}{|c}{MNAR2} \\
$\rho$ & Methods & IA & EA  &  PA \\
\hline
 0.25    & KNN & 1.06(0.070) & 0.82(0.061) & 0.76(0.201) \\
         & RF & 1.02(0.063) & 0.71(0.048) & 0.63(0.173) \\
         & MI & 1.55(0.105) & 0.73(0.042) & 0.67(0.158) \\
         & RI & 1.14(0.100) & 0.78(0.514) & 0.99(1.212) \\
         & FIME & - & 0.64(0.063) & - \\
         & SSI$_1$ & 0.85(0.037) & 0.60(0.034) & 0.62(0.205) \\
         & SSI$_2$ & 0.85(0.038) & 0.60(0.057) & 0.59(0.192) \\
         \hline
 0.5     & KNN & 0.85(0.062) & 0.79(0.046) & 0.64(0.227) \\
         & RF & 0.77(0.046) & 0.74(0.042) & 0.53(0.210) \\
         & MI & 1.19(0.093) & 0.75(0.039) & 0.57(0.196) \\
         & RI & 0.89(0.085) & 0.82(0.499) & 0.80(1.194) \\
         & FIME & - & 0.70(0.061) & - \\
         & SSI$_1$ & 0.71(0.035) & 0.66(0.035) & 0.54(0.235) \\
         & SSI$_2$ & 0.72(0.038) & 0.67(0.050) & 0.51(0.216) \\
         \hline
 0.75    & KNN & 0.55(0.032) & 0.078(0.042) & 0.52(0.254) \\
         & RF & 0.46(0.027) & 0.75(0.039) & 0.46(0.237) \\
         & MI & 0.72(0.063) & 0.76(0.036) & 0.49(0.227) \\
         & RI & 0.59(0.085) & 1.08(2.136) & 0.81(1.440) \\
         & FIME & - & 0.74(0.060) & - \\
         & SSI$_1$ & 0.47(0.025) & 0.71(0.046) & 0.50(0.262) \\
         & SSI$_2$ & 0.49(0.030) & 0.71(0.043) & 0.46(0.237) \\
         \hline
\end{tabular}
\end{center}
\end{table}
\end{landscape}

\begin{landscape}
\begin{table}[htbp!]
\begin{center}
\caption
{The simulation results of IA, EA and PA under the
three coefficients of determination ($R^2$=0.3, 0.6 and 0.9) and missing mechanism MNAR3,
and they are obtained by averaging over three correlations ($\rho=0.25$, 0.5 and 0.75) and three sample sizes ($n=500$, 1,000 and 2,000) for Setting V.
The value in parentheses is the averaged standard error.
Note that FIME is only applicable for parameter estimation, hence, we display its IA and PA as ``-".}
\vspace{0.28 cm}
\begin{tabular}{c|c|ccc}
\hline
 &  &  \multicolumn{3}{|c}{MNAR3} \\
$R^2$ & Methods & IA & EA  &  PA \\
\hline
0.3      & KNN & 1.07(0.325) & 0.89(0.072) & 0.95(0.106) \\
         & RF & 0.91(0.312)	 & 0.77(0.037) & 0.80(0.093) \\
         & MI & 1.42(0.392)  & 0.79(0.026) & 0.82(0.095) \\
         & RI & 0.93(0.316)	 & 1.37(1.131) & 2.08(1.895) \\
         & FIME & - & 0.74(0.06) & - \\
         & SSI$_1$ & 0.87(0.223) & 0.68(0.066) & 0.82(0.093) \\
         & SSI$_2$ & 0.88(0.228) & 0.73(0.047) & 0.80(0.087) \\
         \hline
0.6      & KNN & 1.11(0.317) & 0.82(0.084) & 0.83(0.112) \\
         & RF & 0.92(0.311)	& 0.72(0.042) &	0.58(0.114) \\
         & MI & 1.42(0.392)	& 0.74(0.026) &	0.62(0.119) \\
         & RI & 0.92(0.309)	& 1.31(1.191) &	1.96(2.165) \\
         & FIME & - & 0.67(0.061) & - \\
         & SSI$_1$ & 0.85(0.224) & 0.64(0.063) & 0.60(0.097) \\
         & SSI$_2$ & 0.87(0.221) & 0.66(0.053) & 0.57(0.096) \\
         \hline
0.9      & KNN & 1.01(0.318) & 0.76(0.085) & 0.54(0.143) \\
         & RF & 0.92(0.313)	& 0.67(0.047) &	0.36(0.152) \\
         & MI & 1.42(0.392)	& 0.70(0.027) &	0.42(0.161) \\
         & RI & 0.92(0.311)	& 0.97(0.753) &	1.22(1.686)\\
         & FIME & - & 0.63(0.064) & - \\
         & SSI$_1$ & 0.84(0.228) & 0.62(0.058) & 0.36(0.124) \\
         & SSI$_2$ & 0.85(0.227) & 0.62(0.059) & 0.34(0.122) \\
         \hline
\end{tabular}
\end{center}
\end{table}
\end{landscape}

\begin{landscape}
\begin{table}[htbp!]
\begin{center}
\caption
{The simulation results of IA, EA and PA under the three correlations ($\rho=0.25$, 0.5 and 0.75) and missing mechanism MNAR3, and they are obtained by averaging over the
three coefficients of determination ($R^2$=0.3, 0.6 and 0.9) and three sample sizes ($n=500$, 1,000 and 2,000) for Setting V. The value in parentheses is the averaged standard error.
Note that FIME is only applicable for parameter estimation; hence, we display its IA and PA as ``-".}
\vspace{0.28 cm}
\begin{tabular}{c|c|ccc}
\hline
 &  &  \multicolumn{3}{|c}{MNAR3} \\
$\rho$ & Methods & IA & EA  &  PA \\
\hline
0.25      & KNN & 1.32(0.101) & 0.77(0.065) & 0.82(0.180)\\
         & RF & 1.28(0.076)	 & 0.68(0.056) & 0.70(0.151) \\
         & MI & 1.84(0.102)	 & 0.72(0.045) & 0.74(0.138) \\
         & RI & 1.30(0.086)	 & 1.13(0.948) & 2.06(2.161) \\
         & FIME & - & 0.61(0.065) & - \\
         & SSI$_1$ & 1.10(0.048) & 0.58(0.043) & 0.70(0.174) \\
         & SSI$_2$ & 1.11(0.051) & 0.61(0.062) & 0.67(0.158) \\
         \hline
0.5      & KNN & 1.27(0.082) & 0.82(0.071) & 0.78(0.212) \\
         & RF & 0.93(0.065) & 0.72(0.045) &	0.57(0.201) \\
         & MI & 1.50(0.102)	& 0.74(0.039) &	0.61(0.182) \\
         & RI & 0.99(0.096)	& 1.08(0.718) &	1.54(1.625) \\
         & FIME & - & 0.68(0.057)& - \\
         & SSI$_1$ & 0.90(0.046) & 0.65(0.034) & 0.59(0.215) \\
         & SSI$_2$ & 0.91(0.047) & 0.67(0.046) & 0.56(0.197) \\
         \hline
0.75     & KNN & 0.75(0.065) & 0.94(0.068) & 0.66(0.427) \\
         & RF & 0.53(0.034)	& 0.76(0.037) &	0.47(0.232) \\
         & MI & 0.92(0.078)	& 0.76(0.034) &	0.50(0.223) \\
         & RI & 0.57(0.077)	& 1.38(1.312) &	1.64(2.026)\\
         & FIME & - & 0.74(0.045) & - \\
         & SSI$_1$ & 0.56(0.033) & 0.72(0.038) & 0.49(0.244) \\
         & SSI$_2$ & 0.58(0.036) & 0.73(0.041) & 0.46(0.225) \\
         \hline
\end{tabular}
\end{center}
\end{table}
\end{landscape}

\begin{landscape}
\begin{table}[htbp!]
\begin{center}
\caption
{The simulation results of IA, EA and PA under the
three coefficients of determination ($R^2$=0.3, 0.6 and 0.9) and three missing mechanisms (MCAR, MAR, and MNAR),
and they are obtained by averaging over three correlations ($\rho=0.25$, 0.5 and 0.75) and three sample sizes ($n=500$, 1,000 and 2,000) for random missing pattern of Setting IV.
The value in parentheses is the averaged standard error.
Note that FIME is only applicable for parameter estimation, hence, we display its IA and PA as ``-".}
\vspace{0.28 cm}
\begin{tabular}{c|c|ccc|ccc|ccc}
\hline
 &  &  \multicolumn{3}{|c}{MCAR} & \multicolumn{3}{|c}{MAR} & \multicolumn{3}{|c}{MNAR} \\
$R^2$ & Methods & IA & EA  &  PA & IA & EA  &  PA & IA & EA  &  PA \\
\hline
0.3      & KNN & - & - & - & - & - & - & - & - & - \\
         & RF & 0.70(0.229) & 0.77(0.028) & 0.76(0.078) & 0.72(0.219) & 0.79(0.027) & 0.77(0.074) & 0.74(0.184) & 0.79(0.062) & 0.84(0.133) \\
         & MI & 1.08(0.362) & 0.78(0.024) & 0.78(0.081) & 1.11(0.351) & 0.78(0.024) & 0.78(0.079) & 1.21(0.332) & 0.79(0.023) & 0.81(0.085) \\
         & RI & 0.73(0.249) & 1.15(0.762) & 1.71(1.649) & 0.75(0.247) & 0.99(0.548) & 1.37(1.407) & 0.61(0.164) & 0.75(0.029) & 0.77(0.074) \\
         & FIME & - & 0.75(0.039) & - & - & 0.76(0.039) & - & - & 0.73(0.037) & - \\
         & SSI$_1$ & 0.66(0.172) & 0.71(0.058) & 0.79(0.079) & 0.68(0.163) & 0.71(0.061) & 0.80(0.076) & 0.70(0.185) & 0.74(0.096) & 0.62(0.101) \\
         & SSI$_2$ & 0.67(0.172) & 0.73(0.046) & 0.76(0.075) & 0.68(0.161) & 0.73(0.048) & 0.76(0.071) & 0.80(0.135) & 0.70(0.053) & 0.53(0.078) \\
         \hline
 0.6     & KNN & - & - & - & - & - & - & - & - & - \\
         & RF & 0.69(0.227) & 0.72(0.030) & 0.50(0.074) & 0.72(0.218) & 0.72(0.028) & 0.51(0.076) & 0.74(0.198) & 0.73(0.044) & 0.58(0.114) \\
         & MI & 1.08(0.363) & 0.73(0.024) & 0.54(0.087) & 1.11(0.346) & 0.73(0.024) & 0.54(0.087) & 1.19(0.337) & 0.74(0.022) & 0.59(0.100) \\
         & RI & 0.72(0.254) & 1.08(0.960) & 1.37(1.683) & 0.76(0.243) & 0.87(0.519) & 0.93(1.157) & 0.61(0.169) & 0.69(0.034) & 0.52(0.076) \\
         & FIME & - & 0.69(0.041) & - & - & 0.69(0.041) & - & - & 0.69(0.033) & - \\
         & SSI$_1$ & 0.66(0.172) & 0.66(0.049) & 0.50(0.062) & 0.67(0.160) & 0.66(0.051) & 0.51(0.064) & 0.74(0.157) & 0.65(0.040) & 0.38(0.068) \\
         & SSI$_2$ & 0.67(0.163) & 0.66(0.048) & 0.49(0.063) & 0.68(0.153) & 0.66(0.051) & 0.50(0.065) & 0.74(0.148) & 0.65(0.041) & 0.38(0.068) \\
         \hline
 0.9     & KNN & - & - & - & - & - & - & - & - & - \\
         & RF & 0.69(0.225) & 0.68(0.030) & 0.24(0.087) & 0.72(0.217) & 0.68(0.030) & 0.25(0.088) & 0.72(0.210) & 0.68(0.031) & 0.29(0.095) \\
         & KNN & - & - & - & - & - & - & - & - & - \\
         & MI & 1.08(0.360) & 0.69(0.023) & 0.29(0.109) & 1.11(0.347) & 0.69(0.024) & 0.30(0.107) & 1.14(0.341) & 0.70(0.024) & 0.34(0.111) \\
         & RI & 0.73(0.252) & 0.92(0.767) & 0.86(1.526) & 0.76(0.245) & 0.77(0.391) & 0.54(1.016) & 0.63(0.216) & 0.65(0.036) & 0.25(0.080) \\
         & FIME & - & 0.64(0.043) & - & - & 0.65(0.045) & - & - & 0.65(0.038) & - \\
         & SSI$_1$ & 0.65(0.173) & 0.64(0.039) & 0.23(0.074) & 0.67(0.163) & 0.64(0.042) & 0.23(0.070) & 0.69(0.162) & 0.63(0.037) & 0.24(0.084) \\
         & SSI$_2$ & 0.65(0.167) & 0.62(0.052) & 0.22(0.067) & 0.67(0.164) & 0.63(0.060) & 0.23(0.067) & 0.68(0.155) & 0.63(0.042) & 0.23(0.073) \\
         \hline
\end{tabular}
\end{center}
\end{table}
\end{landscape}

\begin{landscape}
\begin{table}[htbp!]
\begin{center}
\caption
{The simulation results of IA, EA and PA under the three correlations ($\rho=0.25$, 0.5 and 0.75)
and three missing mechanisms (MCAR, MAR, and MNAR), and they are obtained by averaging over the
three coefficients of determination ($R^2$=0.3, 0.6 and 0.9) and three sample sizes ($n=500$, 1,000 and 2,000) for random missing patter of
Setting IV. The value in parentheses is the averaged standard error.
Note that FIME is only applicable for parameter estimation; hence, we display its IA and PA as ``-".}
\vspace{0.28 cm}
\begin{tabular}{c|c|ccc|ccc|ccc}
\hline
 &  &  \multicolumn{3}{|c}{MCAR} & \multicolumn{3}{|c}{MAR} & \multicolumn{3}{|c}{MNAR} \\
$\rho$ & Methods & IA & EA  &  PA & IA & EA  &  PA & IA & EA  &  PA \\
\hline
 0.25    & KNN & - & - & - & - & - & - & - & - & - \\
         & RF & 0.96(0.051) & 0.69(0.049) & 0.57(0.192) & 0.97(0.052) & 0.70(0.047) & 0.58(0.194) & 0.95(0.056) & 0.71(0.068) & 0.65(0.237) \\
         & MI & 1.51(0.091) & 0.71(0.044) & 0.62(0.173) & 1.51(0.092) & 0.71(0.044) & 0.63(0.176) & 1.55(0.097) & 0.72(0.047) & 0.67(0.177) \\
         & RI & 1.03(0.069) & 0.90(0.502) & 1.37(1.624) & 1.04(0.079) & 0.80(0.426) & 1.05(1.376) & 0.82(0.079) & 0.65(0.048) & 0.58(0.197) \\
         & FIME & - & 0.65(0.058) & - & - & 0.65(0.056) & - & - & 0.64(0.040) & - \\
         & SSI$_1$ & 0.85(0.041) & 0.61(0.034) & 0.56(0.209) & 0.85(0.041) & 0.61(0.034) & 0.57(0.215) & 0.89(0.049) & 0.62(0.037) & 0.44(0.108) \\
         & SSI$_2$ & 0.85(0.041) & 0.61(0.054) & 0.54(0.205) & 0.85(0.040) & 0.61(0.057) & 0.55(0.205) & 0.89(0.051) & 0.61(0.042) & 0.43(0.112) \\
         \hline
  0.5    & KNN & - & - & - & - & - & - & - & - & - \\
         & RF & 0.70(0.035) & 0.73(0.041) & 0.49(0.224) & 0.73(0.040) & 0.73(0.040) & 0.50(0.222) & 0.76(0.054) & 0.74(0.062) & 0.56(0.247) \\
         & MI & 1.10(0.066) & 0.73(0.040) & 0.52(0.212) & 1.13(0.072) & 0.74(0.039) & 0.53(0.210) & 1.23(0.094) & 0.75(0.041) & 0.58(0.209) \\
         & RI & 0.75(0.053) & 1.03(0.729) & 1.32(1.773) & 0.77(0.061) & 0.87(0.467) & 0.94(1.249) & 0.64(0.045) & 0.70(0.043) & 0.51(0.222) \\
         & FIME & - & 0.70(0.049) & - & - & 0.70(0.047) & - & - & 0.69(0.035) & - \\
         & SSI$_1$ & 0.69(0.034) & 0.67(0.036) & 0.50(0.238) & 0.70(0.035) & 0.68(0.037) & 0.51(0.239) & 0.74(0.047) & 0.69(0.076) & 0.43(0.192) \\
         & SSI$_2$ & 0.69(0.035) & 0.67(0.047) & 0.48(0.227) & 0.70(0.034) & 0.68(0.044) & 0.49(0.225) & 0.78(0.074) & 0.67(0.040) & 0.38(0.139) \\
          \hline
  0.75   & KNN & - & - & - & - & - & - & - & - & - \\
         & RF & 0.42(0.020) & 0.75(0.039) & 0.44(0.242) & 0.45(0.026) & 0.76(0.037) & 0.44(0.239) & 0.49(0.045) & 0.76(0.053) & 0.49(0.255) \\
         & MI & 0.64(0.045) & 0.75(0.037) & 0.46(0.238) & 0.68(0.046) & 0.76(0.036) & 0.46(0.233) & 0.76(0.065) & 0.76(0.037) & 0.49(0.232) \\
         & RI & 0.43(0.032) & 1.20(1.118) & 1.20(1.561) & 0.46(0.037) & 0.95(0.577) & 0.84(1.098) & 0.39(0.032) & 0.73(0.039) & 0.46(0.238) \\
         & FIME & - & 0.73(0.046) & - & - & 0.74(0.043) & - & - & 0.73(0.037) & - \\
         & SSI$_1$ & 0.43(0.023) & 0.72(0.043) & 0.46(0.257) & 0.47(0.027) & 0.72(0.042) & 0.47(0.256) & 0.50(0.041) & 0.72(0.077) & 0.37(0.212) \\
         & SSI$_2$ & 0.45(0.028) & 0.72(0.041) & 0.44(0.242) & 0.47(0.031) & 0.73(0.040) & 0.44(0.237) & 0.56(0.074) & 0.70(0.042) & 0.33(0.156) \\
         \hline
\end{tabular}
\end{center}
\end{table}
\end{landscape}

\begin{landscape}
\begin{table}[htbp!]
\caption
{The simulation results of IA, EA and PA under the
three coefficients of determination ($R^2$=0.3, 0.6 and 0.9) and three missing mechanisms (MCAR, MAR, and MNAR), and they are
 obtained by averaging over three correlations ($\rho=0.25$, 0.5 and 0.75) and three sample sizes ($n=500$, 1,000 and 2,000).
The value in parentheses is the averaged standard error.}
\begin{center}
\vspace{0.18 cm}
\begin{tabular}{c|c|ccc|ccc|ccc}
\hline
 &  &  \multicolumn{3}{|c}{MCAR} & \multicolumn{3}{|c}{MAR} & \multicolumn{3}{|c}{MNAR} \\
$R^2$ & Methods & IA & EA  &  PA & IA & EA  &  PA & IA & EA  &  PA \\
\hline
0.3 & SSI$_1$ & 0.68(0.157) & 0.70(0.069) & 0.83(0.083) & 0.68(0.158)  & 0.70(0.075) & 0.83(0.087)  & 0.68(0.157) & 0.70(0.065) & 0.81(0.087) \\
    & SSI$_2$ & 0.70(0.149) & 0.72(0.046)  & 0.77(0.079) & 0.70(0.149) & 0.72(0.051) & 0.78(0.083) & 0.73(0.148) & 0.72(0.054) & 0.77(0.081)  \\
    & SSSI$_1$ & 0.67(0.160) & 0.69(0.070) & 0.80(0.082) & 0.66(0.159)  & 0.68(0.074) & 0.82(0.085)  & 0.66(0.155) & 0.69(0.062) & 0.80(0.085) \\
    & SSSI$_2$ & 0.70(0.150) & 0.70(0.046)  & 0.76(0.077) & 0.69(0.147) & 0.71(0.050) & 0.78(0.084) & 0.70(0.150) & 0.70(0.055) & 0.75(0.080)  \\
    \hline
0.6 & SSI$_1$ & 0.68(0.158) & 0.65(0.053) & 0.54(0.072) & 0.68(0.158) & 0.65(0.055) & 0.54(0.072) & 0.68(0.159) & 0.65(0.055) & 0.54(0.074) \\
    & SSI$_2$ & 0.68(0.157) & 0.66(0.053) & 0.52(0.075) & 0.68(0.157) & 0.66(0.051) & 0.52(0.072) & 0.69(0.161) & 0.65(0.058) & 0.52(0.076) \\
    & SSSI$_1$ & 0.67(0.160) & 0.64(0.054) & 0.52(0.070) & 0.65(0.160) & 0.63(0.051) & 0.53(0.076) & 0.66(0.160) & 0.64(0.056) & 0.52(0.072) \\
    & SSSI$_2$ & 0.68(0.156) & 0.64(0.057) & 0.51(0.075) & 0.67(0.155) & 0.62(0.050) & 0.50(0.070) & 0.68(0.162) & 0.63(0.057) & 0.51(0.071) \\
    \hline
0.9 & SSI$_1$ & 0.67(0.162)  & 0.63(0.042) & 0.27(0.083) & 0.67(0.162) & 0.63(0.041) & 0.27(0.083) & 0.67(0.161) & 0.63(0.042) & 0.28(0.087) \\
    & SSI$_2$ & 0.68(0.155) & 0.62(0.053) & 0.26(0.079) & 0.68(0.161)  & 0.62(0.053) & 0.27(0.083) & 0.68(0.155) & 0.61(0.057) & 0.27(0.083) \\
    & SSSI$_1$ & 0.66(0.160)  & 0.62(0.043) & 0.25(0.082) & 0.66(0.158) & 0.62(0.041) & 0.25(0.081) & 0.65(0.160) & 0.60(0.040) & 0.26(0.085) \\
    & SSSI$_2$ & 0.65(0.154) & 0.61(0.055) & 0.25(0.080) & 0.66(0.159)  & 0.62(0.052) & 0.25(0.081) & 0.65(0.156) & 0.59(0.050) & 0.26(0.080) \\
\hline
\end{tabular}
\end{center}
\end{table}
\end{landscape}

\begin{landscape}
\begin{table}[htbp!]
\caption
{The simulation results of IA, EA and PA under the three correlations ($\rho=0.25$, 0.5 and 0.75)
and three missing mechanisms (MCAR, MAR, and MNAR), and they are obtained by averaging over the
three coefficients of determination ($R^2$=0.3, 0.6 and 0.9) and three sample sizes ($n=500$, 1,000 and 2,000).
The value in parentheses is the averaged standard error.}
\begin{center}
\vspace{0.18 cm}
\begin{tabular}{c|c|ccc|ccc|ccc}
\hline
 &  &  \multicolumn{3}{|c}{MCAR} & \multicolumn{3}{|c}{MAR} & \multicolumn{3}{|c}{MNAR} \\
$\rho$ & Methods & IA & EA  &  PA & IA & EA  &  PA & IA & EA  &  PA \\
\hline
0.25 & SSI$_1$ & 0.85(0.038) & 0.60(0.038)   & 0.61(0.208) & 0.85(0.037)  & 0.60(0.036) & 0.61(0.211) & 0.85(0.038) & 0.60(0.039) & 0.61(0.203) \\
     & SSI$_2$ & 0.85(0.038) & 0.61(0.058) & 0.58(0.192) & 0.86(0.040)  & 0.61(0.056) & 0.59(0.194) & 0.86(0.039) & 0.60(0.061) & 0.59(0.192) \\
     & SSSI$_1$ & 0.84(0.037) & 0.59(0.040) & 0.60(0.199) & 0.84(0.040)  & 0.58(0.033) & 0.61(0.210)  & 0.84(0.042) & 0.58(0.037) & 0.60(0.206) \\
     & SSSI$_2$ & 0.85(0.037) & 0.60(0.056)  & 0.57(0.187) & 0.86(0.038) & 0.60(0.050) & 0.57(0.193) & 0.85(0.041) & 0.58(0.060) & 0.57(0.185)  \\
    \hline
0.5  & SSI$_1$ & 0.71(0.035) & 0.66(0.039) & 0.54(0.234) & 0.71(0.035) & 0.66(0.038) & 0.54(0.238) & 0.71(0.035) & 0.66(0.039) & 0.54(0.227) \\
     & SSI$_2$ & 0.72(0.037) & 0.67(0.047) & 0.51(0.216)  & 0.71(0.035) & 0.67(0.050)  & 0.51(0.217) & 0.72(0.042)  & 0.67(0.051)  & 0.51(0.215) \\
     & SSSI$_1$ & 0.70(0.034) & 0.65(0.037) & 0.53(0.219) & 0.70(0.041) & 0.64(0.040) & 0.53(0.218) & 0.70(0.034) & 0.64(0.046) & 0.52(0.218) \\
     & SSSI$_2$ & 0.70(0.038) & 0.64(0.046) & 0.51(0.213) & 0.70(0.038) & 0.63(0.042) & 0.50(0.220) & 0.72(0.040) & 0.63(0.052) & 0.51(0.206) \\
    \hline
0.75 & SSI$_1$ & 0.47(0.024) & 0.71(0.040) & 0.49(0.260)  & 0.47(0.027) & 0.72(0.051) & 0.48(0.259) & 0.47(0.025) & 0.71(0.044) & 0.48(0.245) \\
     & SSI$_2$ & 0.49(0.036)  & 0.72(0.041) & 0.46(0.235) & 0.49(0.038)  & 0.72(0.042) & 0.46(0.237)  & 0.51(0.101) & 0.72(0.047) & 0.46(0.234) \\
     & SSSI$_1$ & 0.46(0.024)  & 0.70(0.043) & 0.47(0.262) & 0.47(0.026) & 0.72(0.050) & 0.48(0.238) & 0.47(0.032) & 0.70(0.040) & 0.46(0.213) \\
     & SSSI$_2$ & 0.49(0.040) & 0.71(0.045) & 0.45(0.228) & 0.48(0.036)  & 0.71(0.044) & 0.45(0.231) & 0.50(0.098) & 0.71(0.050) & 0.45(0.220) \\
\hline
\end{tabular}
\end{center}
\end{table}
\end{landscape}